\documentclass[11pt]{article}
\usepackage{fullpage,citesort,psfig,amssymb}
\usepackage[normal,small]{caption}

\def\to{\rightarrow}
\newcommand{\ket}[1]{|#1\rangle}

\def\m@th{\mathsurround=0pt }
\def\leftrightarrowfill{$\m@th \mathord\leftarrow \mkern-6mu \cleaders\hbox{$\mkern-2mu \mathord- \mkern-2mu$}\hfill
 \mkern-6mu \mathord\rightarrow$}
\def\overleftrightarrow#1{\vbox{\ialign{##\crcr
     \leftrightarrowfill\crcr\noalign{\kern-1pt\nointerlineskip}
     $\hfil\displaystyle{#1}\hfil$\crcr}}}

\begin{document}

\renewcommand{\thefootnote}{\fnsymbol{footnote}}

\begin{titlepage}
\begin{flushright}
UFIFT-HEP-00-17\\
hep-ph/0007019
\end{flushright}

\vskip 2.5cm

\begin{center}
\begin{Large}
{\bf Analyzing the 't Hooft Model on an $x^+$-$p^+$
Lattice\footnote{This work was supported in part by the Department
of Energy under Grant No. DE-FG02-97ER-41029.}
}
\end{Large}

\vskip 2.cm

{\large 
Joel S. Rozowsky\footnote{E-mail  address: {\tt rozowsky@phys.ufl.edu}} 
and Charles B. Thorn\footnote{E-mail  address: {\tt thorn@phys.ufl.edu}}}

\vskip 0.5cm

{\it Institute for Fundamental Theory\\
Department of Physics, University of Florida,
Gainesville, FL 32611}

(\today)

\vskip 1.0cm
\end{center}

\begin{abstract}
We study the 't Hooft model (large $N_c$
QCD in 2 space-time dimensions) 
using an improved approach to digitizing the
sum of gauge theory Feynman diagrams based on light-cone gauge
$A^+=0$ and  discretized $p^+$ and $ix^+$. Our purpose
is to test the new formalism in a solvable case, with the hope 
to gain some insight into how it might be usefully
applied to the physically interesting case of 
$4$ dimensional QCD.
\end{abstract}

\vfill
\end{titlepage}

\setcounter{equation}0
\renewcommand{\theequation}{\thesection.\arabic{equation}}
\section{Introduction}
\label{sec1}
Last year, with Bering we proposed~\cite{beringrt} a new method
to digitize the sum of planar diagrams selected by 't Hooft's
$N_c\to\infty$ limit of $SU(N_c)$ gauge theories~\cite{thooftlargen}. The
proposal, based on the light-cone or infinite momentum
frame description of the dynamics, 
involved discretization of both the $p^+$ carried by
each line of the diagram and the propagation time $\tau=ix^+$, 
as in~\cite{thornfishnet,gilesmt,browergt}. But the
main advantage of the new version
was a coherent prescription for resolving most
of the ambiguities due to $p^+=0$ divergences that typically
plague the light-cone description.

We hope that our formalism will eventually 
allow an improved understanding of QCD in 4 dimensional space-time.
But in this article, we merely wish to test the proposal in
the context of the well-understood case of large
$N_c$ gauge theories in two space-time
dimensions, namely the 't Hooft model~\cite{thooftmodel}. Our purpose
is not to unearth new aspects of the model, but rather to
see how  its well known properties can be obtained from our
new discretization.

The physical content of the 't Hooft model boils down to an
integral equation, essentially a Bethe-Salpeter equation~\cite{bethes},
that determines the mass spectrum of $q\bar q$ mesons. The
reason the limit $N_c\to\infty$ reduces to ladder diagrams
(albeit with self-energy corrected quark propagators),
is that the 2 dimensional gluon is not dynamical (there are
no transverse polarizations). Thus, as with any axial gauge,
the light-cone gauge $A_-=0$ eliminates all gluon self-interactions, so
$A_+$ can be integrated out  inducing
an instantaneous Coulomb potential.
But the 't Hooft limit $N_c\to\infty$ further eliminates all
quark loops and all  non-planar diagrams, leaving only the
planar self energy corrections to the quark propagator, and
the ladder bare gluon exchanges (Coulomb interaction) 
between quark anti-quark lines
in the singlet $q\bar q$ channel. In light-cone parameters
the Bethe-Salpeter equation summing these ladder $q\bar q$
diagrams simplifies to the single variable 't Hooft
integral equation~\cite{thooftmodel}.
\begin{eqnarray}
{\cal M}^2\varphi(x) = \left({1\over x}
+{1\over1-x}\right)\mu^2\varphi(x) 
-{g_s^2N_c\over2\pi}P\int_0^1 dy {\varphi(y)-\varphi(x)\over (y-x)^2},
\label{thoofteq}
\end{eqnarray}
where the integral is understood to be evaluated by the principal value
prescription. The variable $x$ 
is the fraction carried by the quark of the  total $P^+$ of the system 
(the anti-quark carries $P^+$ fraction $1-x$). Also ${\cal M}$ is
the mass of the meson bound state and $\varphi$ satisfies the
boundary conditions, $\varphi(0)=\varphi(1)=0$.

Since the new formalism discretizes 
$\tau\equiv ix^+=k a$ in addition to $p^+=lm$,
the corresponding simplifications lead to an
equation that is not a straightforward discretization of this
integral equation. In particular, the continuum limit can
be taken in different ways depending on the ratio $T_0=m/a$
(which would be infinite for continuous $\tau$), and we want
to explore to what extent these different continuum limits lead
to the same physics. We shall find that some care must
be taken with the setup of the discrete $\tau$ dynamics in
order for this to be true. Indeed, a numerical study shows that 
the most simple-minded treatment leads to a  ground state
that becomes unstable at moderate 't Hooft coupling even
with relatively small $P^+/m\equiv M$ unless 
the ratio $a/m=1/T_0$ is
tuned to be sufficiently small (perhaps infinitesimal for large $M$).
If this feature were robust, it would cast doubt on any potential 
utility of the discretization of $\tau$.

To overcome this difficulty, we find it necessary to veto
some of the ``densest'' discretized Feynman
diagrams: a quark must be forbidden
to emit 2 gluons at immediately successive time steps,
with a similar veto on two successive absorptions. With
this simple veto (which is prescribed locally in time),
we shall show that the continuum limit reduces to the
't Hooft model provided only that the total $P^+$
of the $q\bar q$ system is  large compared to the 
discretization unit $m$. In particular it is
not necessary that the ratio $T_0=m/a$ be large. Keeping
$T_0$ finite in the continuum limit leads to the
't Hooft equation with a non-trivial renormalization of the
coupling. Because of this effect, it turns out that
the effective (renormalized) coupling is small for both large and small
bare coupling, reminiscent of strong/weak coupling duality.    
The strong coupling limit favors the densest diagrams, so
vetoing some of the densest ones has a dramatic effect
on the strong coupling behavior of the theory. This possibility
was anticipated and discussed in~\cite{beringrt} in connection
with the nature of the fishnet diagrams in higher dimensional
space-time.

The rest of the paper is organized as follows. In Section 2 we set
up the discretized 't Hooft model. We analyze it using a
single time-step transfer matrix in Section 3 and using a
Bethe-Salpeter approach in Section 4. In Section 5 we
discuss and implement the veto which allows a satisfactory
continuum limit at fixed $T_0$. Discussion and concluding
remarks are the subject of the final Section.
\setcounter{equation}0
\renewcommand{\theequation}{\thesection.\arabic{equation}}
\section{Discretized 't Hooft Model}
\label{sec2}
The Lagrange density for $SU(N_c)$ gauge fields coupled to quarks in the
fundamental representation  is given by
\begin{eqnarray}
{\cal L} = -{1\over4}{\rm Tr}F^{\mu\nu}F_{\mu\nu}+
\bar{q}\left[i\gamma\cdot(\partial-igA)-\mu_0\right]q,
\end{eqnarray}
where $F_{\mu\nu}=\partial_\mu A_\nu-\partial_\nu A_\mu-ig[A_\mu,A_\nu]$. 
We remind the reader that the normalization of
gauge fields appropriate for matrix fields and
dictated by the gluon kinetic term differs by a
factor $1/\sqrt{2}$ from the more standard one:
$$-{1\over4}\sum_aF_a^{\mu\nu}F_{a\mu\nu}=-{1\over2}{\rm Tr}F_s^{\mu\nu}
F_{s\mu\nu},$$
with $F_s\equiv\sum_a{\lambda_a\over2}F_a$. Thus $A_s=A/\sqrt{2}$,
and we conclude that $g=g_s/\sqrt2$.
In 2 space-time dimensions we choose the representation
of $\gamma$ matrices for which the light-like components are 
\begin{eqnarray}
\gamma^+=\sqrt{2}\left(
\begin{array}{cc}0&1\\0&0\end{array}\right)
\qquad 
\gamma^-=\sqrt{2}\left(\begin{array}{cc}0&0\\1&0\end{array}\right).
\end{eqnarray}
With this choice the field equation for the upper component of
the quark spinor does not involve the ``time'' derivative and
is an equation of constraint relating the upper component, $q_1$, to the
lower component, $q_2$.
Working in light-cone gauge ($A_-=A^+=0$), 
we can eliminate the upper component in favor of the lower component yielding 
the light-cone gauge Lagrange density
\begin{eqnarray}
{\cal L} = +{1\over2}{\rm Tr}(\partial_- A_+)^2 +
i{\psi}^\dagger\left[\partial_+-igA_++{\mu_0^2\over2\partial_-}
\right]\psi,
\end{eqnarray}
where $\psi=2^{1/4}q_2$.   

Our discretization of Feynman diagrams is based on the $x^+$ representation
of each bare propagator
\begin{eqnarray}
D(p^+,x^+) = \int {dp^-\over2\pi} \tilde{D}(p^+,p^-)e^{-ix^+p^-}.
\end{eqnarray}
Performing the $p^-$ integral gives the following Feynman rules for the
continuum theory 
\begin{eqnarray}
D_\psi(p^+,x^+) &=& e^{-ix^+\mu_0^2/2p^+} \to e^{-\tau\mu_0^2/2p^+} \nonumber\\
D_{A} (p^+,x^+) &=& i{\delta(x^+)\over {p^+}^2}\to -{\delta(\tau)\over
{p^+}^2}
\label{ContinuumRules} \\
V_{\psi^\dagger\psi A} &=& ig \to g, \nonumber
\end{eqnarray}
where the arrows indicate the rules to use with imaginary time.

One way to digitize the 't Hooft equation (\ref{thoofteq}) 
is to put the variables $x, y$
on a grid, which amounts to  discrete light-cone 
quantization~\cite{discretelc,thornfishnet}, 
where one discretizes the amount of  
$P^+$ each line of the ladder diagram carries in quanta of $m$
$$
p^+ = lm \qquad l=1,2,3,\ldots.
$$
One can then focus on a state of the system of interest
(in our case a $q\bar q$ system) with total $P^+=Mm$. 
The continuum theory is recovered by taking the combined
limits $m\to0$ and $M\to\infty$ while keeping $P^+=Mm$ fixed.
Following~\cite{thornfishnet,beringrt}, in addition to discretizing 
the $p^+$ of each particle, we
also discretize imaginary light-cone time, $\tau=ix^+=ka$
($k=1,2,3,\ldots$). This discretization (which also serves as an
ultraviolet cutoff) allows the continuum limit to
be reached by keeping $T_0\equiv m/a$ fixed and taking both $m,a\to 0$ and
$M\to\infty$ simultaneously. Actually, since the physics of
the discretized model depends only
on the ratio $m/a$, the continuum limit is nothing but the
large $M$ limit, where $M$ measures the total $P^+$ of the
system state. The conventional continuous time DLCQ approach
(see~\cite{brodskyppreport} and references therein)
corresponds to the special case $T_0\to \infty$. 

\begin{figure}[ht]
\begin{center}
\begin{tabular}{|c|c|c|c|c|c|}
\hline
&&&&&\\[-.3cm]
$
\begin{array}[c]{c}
\psfig{file=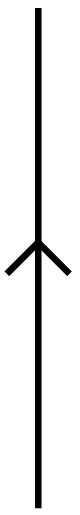,height=0.5in}
\end{array}
$
&
$e^{-k\mu_0^2/2MT_0}$
&
$
\begin{array}[c]{c}
\psfig{file=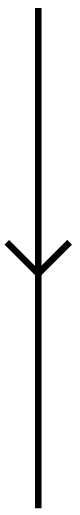,height=0.5in}
\end{array}
$
&
$-e^{-k\mu_0^2/2MT_0}$
&
$
\begin{array}[c]{c}
\psfig{file=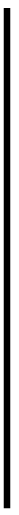,height=0.5in}
\end{array}
$
&
$-f_k{T_0\over M^2}$ \\
\hline
\multicolumn{2}{|c|}{}&&\multicolumn{2}{|c|}{}&\\[-.3cm]
\multicolumn{2}{|c|}{
$
\begin{array}[c]{c}
\psfig{file=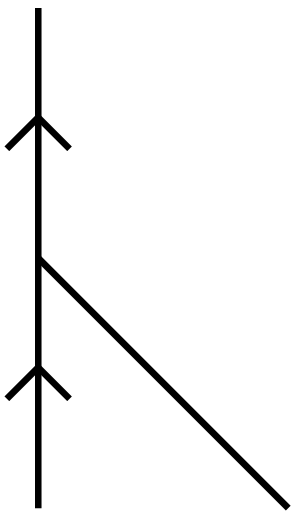,height=0.5in}
\quad
\psfig{file=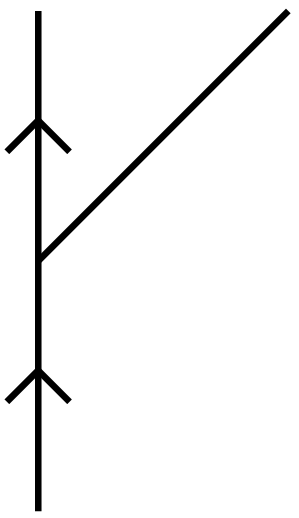,height=0.5in}
\end{array}
$}
&
${g\over T_0\sqrt{2\pi}}$
&
\multicolumn{2}{|c|}{
$
\begin{array}[c]{c}
\psfig{file=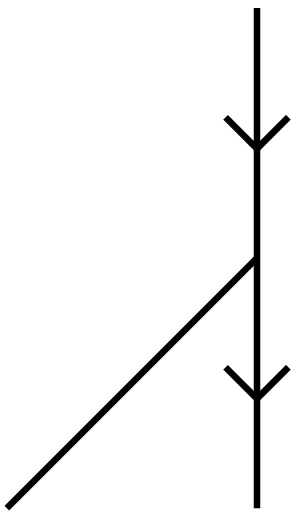,height=0.5in}
\quad
\psfig{file=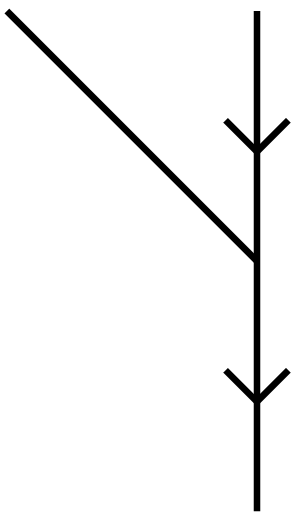,height=0.5in}
\end{array}
$}
&
${g\over T_0\sqrt{2\pi}}$ \\
\hline
\end{tabular}
\end{center}
\caption[]{Feynman rules for the discretized 't Hooft model. Discrete
light-cone time flows up the page.}
\label{FeynmanRules1}
\end{figure}

Discretization of the quark propagator poses no difficulty.
However, for the
instantaneous interaction induced by integrating out $A_+$, 
we allow for ambiguities as in~\cite{beringrt}. The only constraint
is that the discretized propagator become that of
Eq.~\ref{ContinuumRules} in the continuum limit. This allows us 
to spread out the instantaneous interaction away from $\tau=0$ 
(see~\cite{beringrt} for further discussion). Thus the gauge
propagator can be expressed as  
\begin{eqnarray}
D_A (Mm,-ika) = -f_k{T_0\over M^2} \qquad {\rm where} \qquad
\sum_{k=1}^\infty f_k=1.
\label{dummy}
\end{eqnarray}
We require that these arbitrary parameters $f_k$ rapidly vanish with
increasing $k$. 
Using this discretization,
the Feynman rules for the discrete theory are summarized in
Fig.~\ref{FeynmanRules1}.

\begin{figure}[ht]
\begin{center}
\begin{tabular}{|c|c|c|c|c|c|}
\hline
&&&&&\\[-.3cm]
$
\begin{array}[c]{c}
\psfig{file=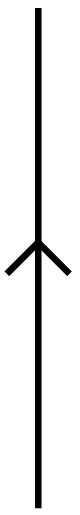,height=0.5in}
\end{array}
$
&
$\alpha^{k/M}$
&
$
\begin{array}[c]{c}
\psfig{file=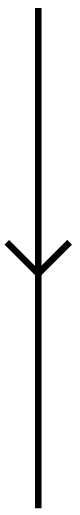,height=0.5in}
\end{array}
$
&
$\alpha^{k/M}$
&
$
\begin{array}[c]{c}
\psfig{file=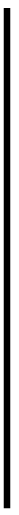,height=0.5in}
\end{array}
$
&
$-{1\over M^2}$ \\
\hline
\multicolumn{2}{|c|}{}&&\multicolumn{2}{|c|}{}&\\[-.3cm]
\multicolumn{2}{|c|}{
$
\begin{array}[c]{c}
\psfig{file=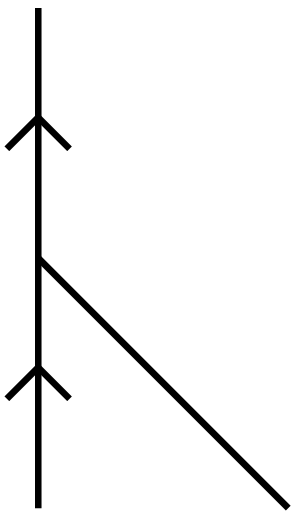,height=0.5in}
\quad
\psfig{file=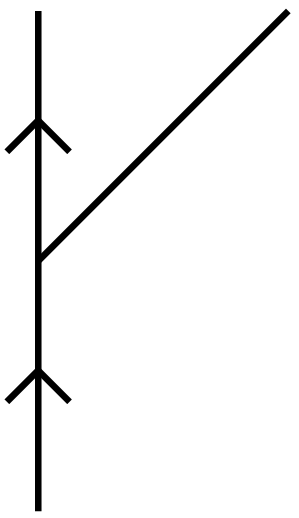,height=0.5in}
\end{array}
$}
&
${\kappa}$
&
\multicolumn{2}{|c|}{
$
\begin{array}[c]{c}
\psfig{file=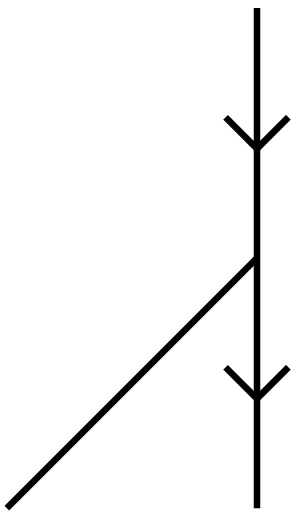,height=0.5in}
\quad
\psfig{file=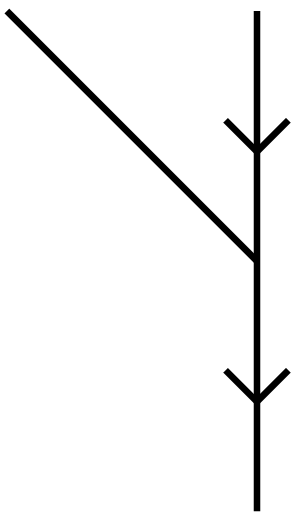,height=0.5in}
\end{array}
$}
&
$-{\kappa}$ \\
\hline
\end{tabular}
\end{center}
\caption[]{Simplified discretized Feynman Rules for 't Hooft model.}
\label{FeynmanRules2}
\end{figure}

For the purposes of this paper we shall not exploit the full generality of
the set of $\{f_k\}$'s. We restrict attention to the simplest version
where the spread out interaction propagates only one unit in
light-cone time, this corresponds to setting $f_1=1, f_{k>1}=0$.
The Feynman rules of Fig.~\ref{FeynmanRules1} can be further 
simplified if we absorb the negative sign from the
anti-quark propagator into the corresponding vertex factor. 
We define new parameters 
\begin{eqnarray}
\alpha\equiv e^{-\mu_0^2/2T_0} \quad{\rm and}\quad 
\kappa\equiv {\sqrt{g^2N_c\over2\pi T_0}}.
\end{eqnarray}
We also recall that in 't Hooft's large $N_c$ limit every additional pair
of cubic vertices in the ladder sum corresponds to a
completed color index loop, which produces a factor $N_c$. Thus we shall
also absorb a factor 
of $\sqrt{N_c}$  into each
vertex. Simply put, all terms in the ladder sum are only dependent
on the 't Hooft coupling $g^2N_c$. The simplified
Feynman rules are presented in Fig.~\ref{FeynmanRules2}. 

\setcounter{equation}0
\renewcommand{\theequation}{\thesection.\arabic{equation}}
\section{Single Time-Step Transfer Matrix}
\label{sec3}
Using the Feynman rules of Fig.~\ref{FeynmanRules2} we can now proceed
to set up a transfer matrix which evolves a singlet $q\bar q$ 
system one step forward
in $x^+$-time. Once the matrix has been determined as a function of
the coupling, $\kappa$, solving the eigenvalue problem will yield 
the bound state energies as functions of coupling. Since the
scalar particle which mediates the Coulomb interaction only lives one
time-step, any state can have at most two intermediate scalars. Thus
for the simplest systems with $P^+/m\equiv M=3,4,5,6$ the number of
states are $3,7,14,25$ (the number of states is $(M-1)(M^2-2M+6)/6$
for general $M$). For illustrative purposes we shall explicitly present the
transfer matrix for $M=4$.
 
For $M=4$ there are 7 states namely:
\begin{eqnarray}
\begin{array}{ccccc}
\ket{3,1} = b^{\dagger}_3d^{\dagger}_1\ket{0}, &&
\ket{2,2} = b^{\dagger}_2d^{\dagger}_2\ket{0}, && 
\ket{1,3} = b^{\dagger}_1d^{\dagger}_3\ket{0}, \\[0.1cm]
\ket{2,1,1} = b^{\dagger}_2a^{\dagger}_1d^{\dagger}_1\ket{0}, &&
\ket{1,2,1} = b^{\dagger}_1a^{\dagger}_2d^{\dagger}_1\ket{0}, && 
\ket{1,1,2} = b^{\dagger}_1a^{\dagger}_1d^{\dagger}_2\ket{0}, 
\\[0.1cm]
&&\ket{1,1,1,1} = b^{\dagger}_1{a^{\dagger}_1}^2d^{\dagger}_1\ket{0}, && 
\end{array}
\label{7states}
\end{eqnarray}
where $b^{\dagger}$, $d^{\dagger}$, and $a^{\dagger}$ are creation
operators for the quark, anti-quark and intermediate gauge particle
states (the subscript on these operators denotes $p^+/m$).  

By construction each of the quark and anti-quark states has at least
one unit of $p^+/m$. The matrix that evolves the system forward in
$x^+$ can be factored into a matrix $A$ that involves only propagators and
a matrix $V$ that involves vertices. Writing the state of the system as a
column vector, $\Upsilon$, with 7 components corresponding to the seven
states in Eq.~\ref{7states}, the transfer matrix equation is
\begin{eqnarray}
t\Upsilon = AV\Upsilon,
\end{eqnarray}
where
\begin{eqnarray}
A&=&{\rm diag}\left[\alpha^{4/3},\alpha,\alpha^{4/3},-\alpha^{3/2}
-\alpha^2/4,-\alpha^{3/2},\alpha^2\right],\\[.5cm]
V&=&
\left(
\begin{array}{ccccccc}
1&0&0&\kappa&\kappa&0&0 \\
0&1&0&-\kappa&0&\kappa&-\kappa^2 \\
0&0&1&0&-\kappa&-\kappa&0 \\
\kappa&-\kappa&0&0&0&-\kappa^2&0 \\
\kappa&0&-\kappa&0&0&0&0 \\
0&\kappa&-\kappa&-\kappa^2&0&0&0 \\
0&-\kappa^2&0&0&0&0&0 
\end{array}
\right),
\end{eqnarray}
and the eigenvalue is $t=e^{-aE}$. Solving this eigenvalue problem will yield
energy eigenvalues as a function of the coupling $\kappa$. Note that 
the matrix $AV$ is not hermitian, and because of the negative diagonal
entries in $A$, the equivalent matrix $\sqrt{A}V\sqrt{A}$ is not
hermitian either. Thus there will, in general be complex eigenvalues $t$.
The best one can hope for is that the lowest lying energy eigenvalues
(highest lying positive real part for $t$) are real. A satisfactory
outcome for the continuum limit $M\to\infty$ would be that the ground
state energy and all the energy values with real parts of order $1/M$ above
the ground state energy are real. Then the complex eigenvalues 
would be strict lattice artifacts.

\begin{figure}[ht]
\centerline{\psfig{file=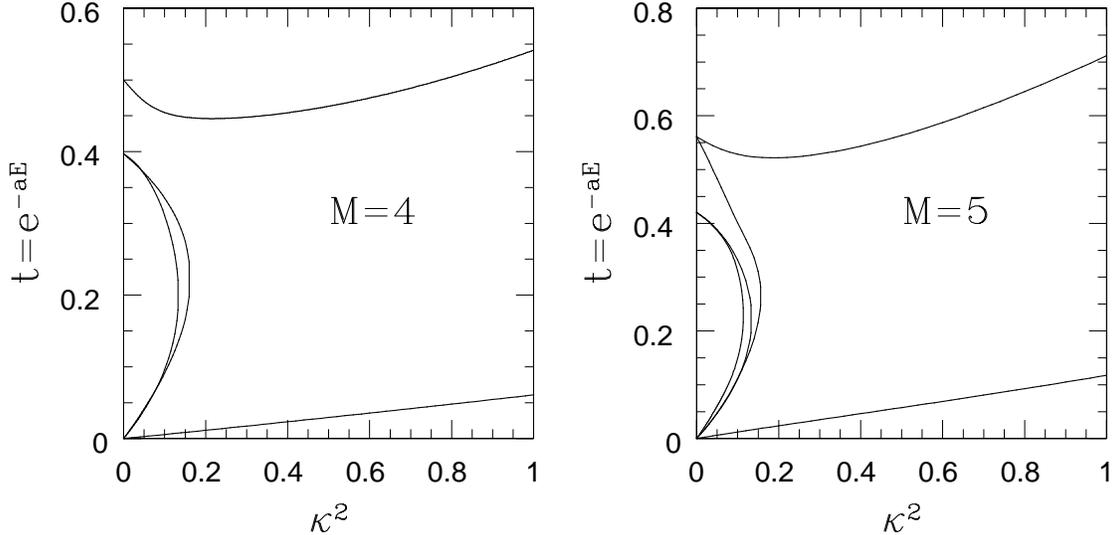,width=6.5in}}
\caption[]{Plot of the real solutions of $t$ as a function of
$\kappa^2$ for the $M=4$ single time-step transfer matrix. It is
convenient to plot $t$ rather than energy since then infinite energy
corresponds to $t=0$. Also note that the lowest lying states are those
with the largest value of $t$.}
\label{m45sts}
\end{figure}

The existence of complex $t$ eigenvalues is already evident at $M=4$
as shown in Fig.~\ref{m45sts}, where we have chosen $\alpha=0.5$
which for definiteness we use in subsequent graphs unless
otherwise indicated.  The ground state (highest
value) of $t$ stays real and positive for all coupling. However the next 
2 excited states stay real only for $\kappa<\kappa_c$ when they
collide with eigenvalues that have emerged from $t=0$ (infinite
energy) after which the eigenvalues become complex conjugate
pairs. The hope is that for increasing $M$ the number of lowest lying
energy levels that remain real all the way to strong coupling should
increase. For $M=4$ analysis shows that the lowest energy eigenvalue
(that of the ground state) stays  well-separated from the 
other states (real and
complex) for all couplings, see Fig.~\ref{m45sts}. 
We also see the eigenvalue solutions (again see Fig.~\ref{m45sts})
which are well behaved at weak coupling can merge with $t=0$ solutions
(solutions which have $t=0$ at zero coupling correspond to infinite
energy lattice artifacts) and become complex. Complex $t$ solutions
are not physical as they correspond to complex energies. This behavior is
generic for our discretization, but as we shall see later, when the
problem has been set up correctly, we can separate the
lowest lying states which survive the continuum
limit from the lattice artifacts.

\begin{figure}[ht]
\centerline{\psfig{file=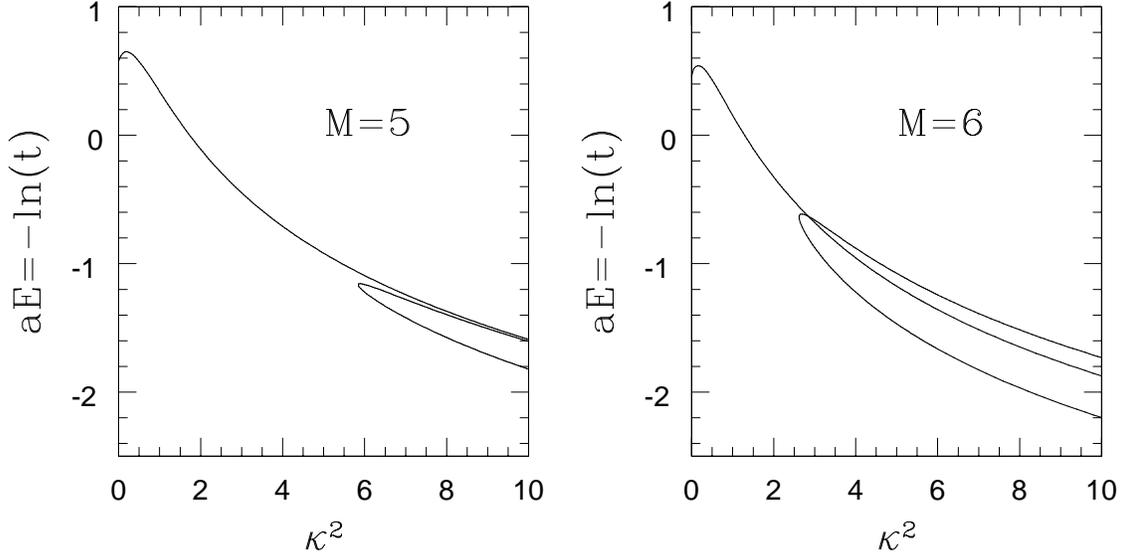,width=6.5in}}
\caption[]{Plot of the lowest energy 
real solutions of $E=-\ln(t)/a$ as a function of
$\kappa^2$ for the $M=5$ and $M=6$ single time-step transfer matrices. We see
the appearance of additional real solutions at lower energies than the
weak coupling ground state for $\kappa>\kappa_c$.}
\label{M56singletimestep}
\end{figure}
However, when one performs a similar analysis for the $M=5$
and $M=6$ systems the lowest eigenstate at weak coupling does not
remain the ground state for all coupling. In both cases a complex
solution at weaker coupling becomes real at larger coupling with a lower
energy than the weak coupling ground state. 
Comparing this behavior for $M=5$ and $M=6$ suggests 
that for increasing $M$ this probably occurs at weaker coupling. Thus for large
$M$ the weak coupling ground state might only be valid for extremely
weak (perhaps only infinitesimal) coupling.  

Conventional continuous time DLCQ corresponds in our discretization to
$\kappa^2\rightarrow0$ since then $T_0\to\infty$. 
In order for our light-cone time
discretization to be useful, the solution should work for all coupling
(corresponding to all values of $T_0$). Here, in this single time-step
analysis, we see that our most naive discretization does not 
satisfy this requirement. We
shall have to modify the discretization in order to fix this.

Since the continuum limit requires
$M\rightarrow\infty$ the single time-step analysis is also inefficient because
the rank of the matrix to diagonalize is of ${\cal O}(M^3)$. However, as we
shall show in the following section, writing the ladder equation in
the form of a Bethe-Salpeter equation (exchange-to-exchange rather
than single time-step) will reduce the complexity of the eigenvalue
problem to a matrix of rank of ${\cal O}(M)$.

\setcounter{equation}0
\renewcommand{\theequation}{\thesection.\arabic{equation}}
\section{Bethe-Salpeter equation}
\label{sec4}
A more efficient way to solve the discretized 't Hooft model is by
setting up a Bethe-Salpeter equation~\cite{bethes}. Instead of a
matrix equation that evolves the $q\bar{q}$ system one step forward in time,
we can write down a system of equations (also a matrix equation)
which evolves the system exchange to exchange. The simplification is
that the intermediate state involves two (dressed)
particles ($M-1$
possible states for general $M$) rather than two, three, and four
bare particles as in
the case of the single time-step transfer matrix. The
trade-off is that the equations become more complicated because of the
dressed propagators.

In order to set up the Bethe-Salpeter equation it is necessary to
work out the dressed quark propagator. In the context of this
discretization the dressed quark propagator is just the sum of all
possible iterated bubbles. There is no room
for nested bubbles because $f_{k>1}=0$. 
While the bubbles extend only one time-step
in $x^+$, we must still allow for all 
possible $P^+$ routings through each bubble.
\begin{figure}[ht]
\centerline{\psfig{file=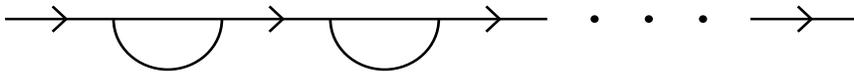,width=4.5in}}
\caption[]{Iterated bubbles which contribute to the quark
propagator.}
\label{DressedProp}
\end{figure}

The energy representation of 
the bare quark propagator carrying 
$p^+/m=l$ (without bubbles), obtained by multiplying by $u^k$ and summing over
all $k>0$, is given by
\begin{eqnarray}
D_q(l) = \sum_{k=1}^{\infty} \left(u\alpha^{1/l}\right)^k 
= {u\alpha^{1/l}\over1-u\alpha^{1/l}},
\label{Prop}
\end{eqnarray}
where $u\equiv1/t=e^{aE}$. The contribution of a single bubble is
\begin{eqnarray}
-\kappa^2\Sigma_l\equiv-\kappa^2\sum_{r=1}^{l-1}{1\over r^2}\alpha^{1/(l-r)}.
\label{Bubble}
\end{eqnarray}
The full propagator is given by iterations of Eq.~\ref{Prop} and
Eq.~\ref{Bubble} as displayed in Fig.~\ref{DressedProp} 
\begin{eqnarray}
D^{\rm full}_q(l)
=D_q(l)\sum_{k=0}^{\infty}\left(-u\kappa^2\Sigma_lD_q(l)\right)^k
={t\alpha^{1/l}\over t^2-t\alpha^{1/l}+\alpha^{1/l}\kappa^2\Sigma_l}.
\end{eqnarray}
The denominator of the full propagator can be factored in two roots so
that
\begin{eqnarray}
D^{\rm full}_q(l)
={t\alpha^{1/l}\over (t-t_+)(t-t_-)},
\end{eqnarray}
where
\begin{eqnarray}
t_{\pm}={\alpha^{1/l}\over2}
\left[1\pm\sqrt{1-4\alpha^{-1/l}\kappa^2\Sigma_l}\right].
\label{Roots}
\end{eqnarray}
We can now partial fraction the full propagator
\begin{eqnarray}
D^{\rm full}_q(l)
&=&{\alpha^{1/l}\over (t_+-t_-)}
\left[{t_+\over(t-t_+)}-{t_-\over(t-t_-)}\right]
\nonumber\\
&=&{\alpha^{1/l}\over (t_+-t_-)}\sum_{k=1}^{\infty}u^k
\left(t_+^k-t_-^k\right).
\label{PropSum}
\end{eqnarray}
Expressing the full quark propagator as the sum in Eq.~\ref{PropSum}
allows us to read off the time representation of the 
full quark propagator for discrete $\tau=ka$. 

\begin{figure}[ht]
\centerline{\psfig{file=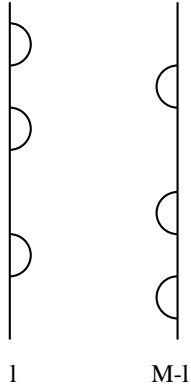,height=2in}}
\caption[]{Here is a section of the ladder sum between two
exchanges. The quark on the left carries $p^+/m=l$ and the anti-quark
on the right carries $p^+/m=M-l$.}
\label{BubblesProp}
\end{figure}

What we really need in order to set up the Bethe-Salpeter equation is a
`propagator' which propagates the $q\bar q$ system, 
including bubbles, between
exchanges between the quark and anti-quark, see Fig.~\ref{BubblesProp}. 
The `propagator' which evolves the system forward between
exchanges is then 
\begin{eqnarray}
{\cal D}_{q\bar{q}}(l)
={\alpha^{M/l(M-l)}\over(t_+-t_-)(s_+-s_-)}\sum_{k=1}^{\infty}u^k
\left(t_+^k-t_-^k\right)\left(s_+^k-s_-^k\right),
\end{eqnarray}
where $s_{\pm}$ are the roots for the anti-quark (obtained
simply by replacing $l$ in Eq.~\ref{Roots} by $M-l$). With some
manipulation this can be simplified to
\begin{eqnarray}
{\cal D}_{q\bar{q}}(l)
={u\alpha^{M/l(M-l)}\left(1-u^2\kappa^4\Sigma_l'\Sigma_{M-l}'\right)
\over \left(1-u^2\kappa^4\Sigma_l'\Sigma_{M-l}'\right)^2
-u\alpha^{M/l(M-l)}\left(1-u\kappa^2\Sigma_{M-l}'\right)
\left(1-u\kappa^2\Sigma_{l}'\right)},
\label{MesonProp}
\end{eqnarray}
where for brevity, we have defined
\begin{eqnarray}
\Sigma_l'\equiv\alpha^{1/(M-l)}\Sigma_l.
\end{eqnarray}

\begin{figure}[ht]
\centerline{\psfig{file=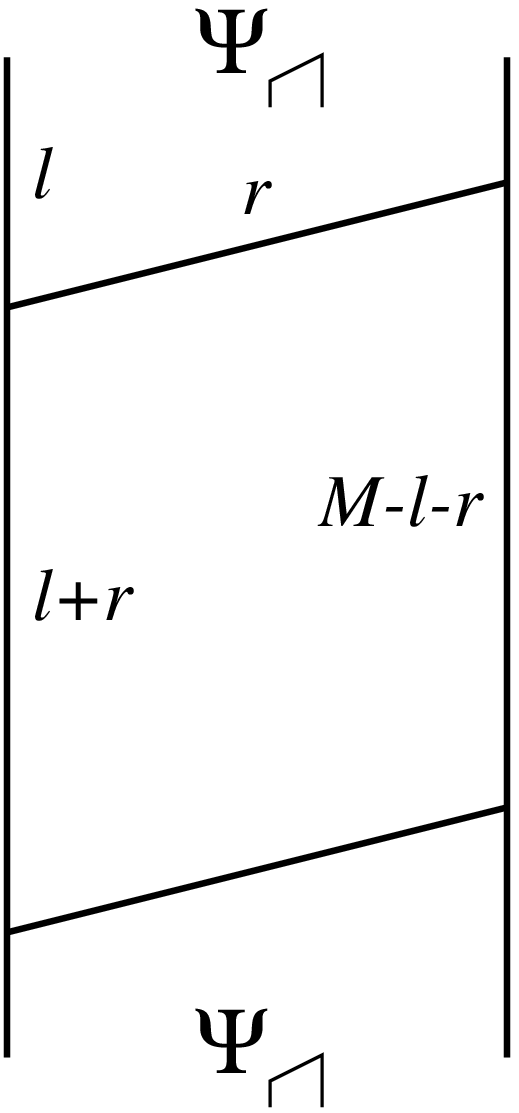,height=2.0in}\hskip 2cm
\psfig{file=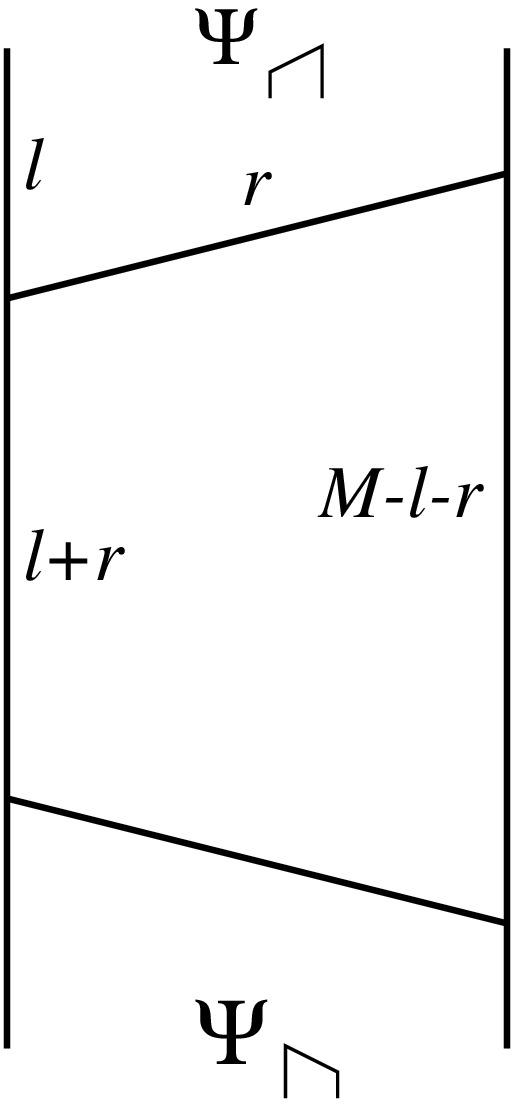,height=2.0in}}
\caption[]{Parallelogram and trapezoid sections of the ladder sum. 
Internal variables label the number of units of $p^+/m$ carried by
each leg. The quark and anti-quark propagators include self-energy
corrections.}
\label{Ladder}
\end{figure}

We can now now set up the Bethe-Salpeter equations
\begin{eqnarray}
\Psi_{\,\psfig{file=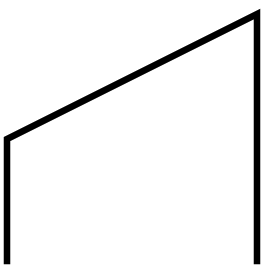,height=.075in}\,}(l)
&=&\sum_{r=1}^{M-l-2}{\kappa^2\over r^2}
{\cal D}_{\,\psfig{file=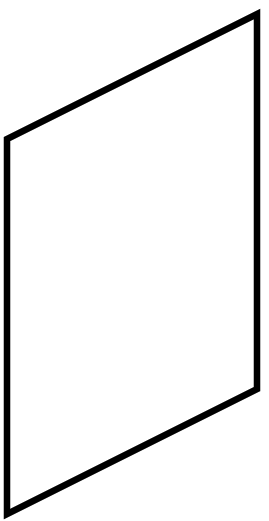,height=.15in}}(l+r)
\Psi_{\,\psfig{file=right.eps,height=.075in}\,}(l+r)
+\sum_{r=1}^{M-l-1}{\kappa^2\over r^2}
{\cal D}_{\,\psfig{file=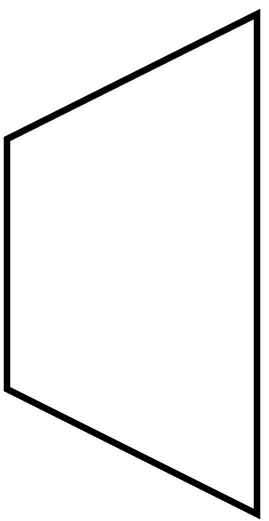,height=.15in}}(l+r)
\Psi_{\,\psfig{file=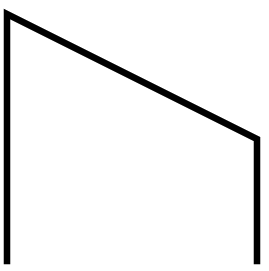,height=.075in}\,}(l+r) 
\nonumber \\
\Psi_{\,\psfig{file=left.eps,height=.075in}\,}(l)
&=&\sum_{r=1}^{l-1}{\kappa^2\over r^2}
{\cal D}_{\,\psfig{file=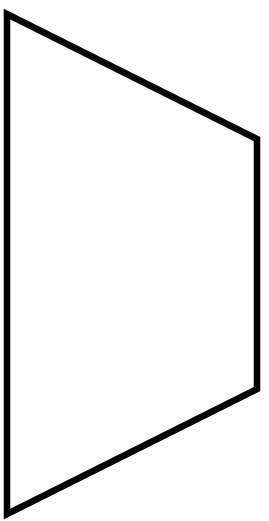,height=.15in}}(l-r)
\Psi_{\,\psfig{file=right.eps,height=.075in}\,}(l-r)
+\sum_{r=1}^{l-2}{\kappa^2\over r^2}
{\cal D}_{\,\psfig{file=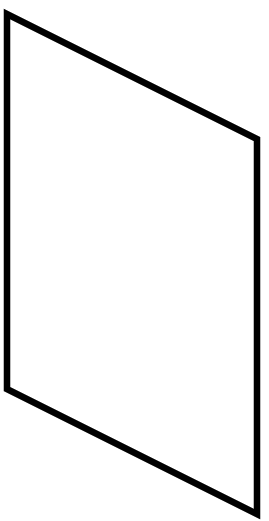,height=.15in}}(l-r)
\Psi_{\,\psfig{file=left.eps,height=.075in}\,}(l-r). 
\label{BetheSalpeterNoVeto}
\end{eqnarray}
$\Psi_{\,\psfig{file=right.eps,height=.075in}\,}$, 
$\Psi_{\,\psfig{file=left.eps,height=.075in}\,}$ 
label two-particle states where the last ladder rung propagated
forward in time from left
to right or right to left, respectively. The first equation is graphically
portrayed in Fig.~\ref{Ladder}.
Since each of the quark/anti-quark propagators must carry a minimum of
one unit of $p^+/m$ there are only $2(M-2)$ possible states
\begin{eqnarray}
\Psi_{\,\psfig{file=right.eps,height=.075in}\,}(l) 
\quad : \quad 1\leq l\leq M-2, \qquad\quad
\Psi_{\,\psfig{file=left.eps,height=.075in}\,}(l) 
\quad : \quad 2\leq l\leq M-1. 
\end{eqnarray}

Eq.~\ref{BetheSalpeterNoVeto} is constructed by evolving the system
from a state just after one exchange in the ladder sum to just
after the next. The various ${\cal D}$'s in
Eq.~\ref{BetheSalpeterNoVeto} correspond to the Feynman diagram
contributions which are either parallelogram or trapezoidal sections
which take a $\Psi_{\,\psfig{file=left.eps,height=.075in}\,}$ or 
$\Psi_{\,\psfig{file=right.eps,height=.075in}\,}$ to a 
$\Psi_{\,\psfig{file=left.eps,height=.075in}\,}$ or
$\Psi_{\,\psfig{file=right.eps,height=.075in}\,}$. 
The parallelogram propagator sections are simply related to
Eq.~\ref{MesonProp}, 
\begin{eqnarray}
{\cal D}_{\,\psfig{file=trap1.eps,height=.15in}}(l)&=&{\cal D}_{q\bar{q}}(l)
\nonumber \\
{\cal D}_{\,\psfig{file=trap2.eps,height=.15in}}(l)&=&{\cal
D}_{q\bar{q}}(M-l)={\cal D}_{q\bar{q}}(l).
\end{eqnarray}
However, the trapezoidal segments must be independently determined 
\begin{eqnarray}
{\cal D}_{\,\psfig{file=par2.eps,height=.15in}}(l)&=&\bar{\cal D}_{q\bar{q}}(l)
\nonumber \\
{\cal D}_{\,\psfig{file=par1.eps,height=.15in}}(l)&=&
\bar{\cal D}_{q\bar{q}}(M-l),
\end{eqnarray}
where
\begin{eqnarray}
\bar{\cal D}_{q\bar{q}}(l)&=&
{\alpha^{M/l(M-l)}\over(t_+-t_-)(s_+-s_-)}\sum_{k=1}^{\infty}u^{k+1}
\left(t_+^k-t_-^k\right)\left(s_+^{k+2}-s_-^{k+2}\right)
\nonumber \\
&=&{u\alpha^{2/(M-l)}\left[
u\alpha^{M/l(M-l)}\left(1-u\kappa^2\Sigma_{M-l}'\right)
-\kappa^2\Sigma_{M-l}'\left(1-u^2\kappa^4\Sigma_l'\Sigma_{M-l}'\right)
\right]
\over \left(1-u^2\kappa^4\Sigma_l'\Sigma_{M-l}'\right)^2
-u\alpha^{M/l(M-l)}\left(1-u\kappa^2\Sigma_{M-l}'\right)
\left(1-u\kappa^2\Sigma_{l}'\right)}.
\end{eqnarray}

In order to solve the matrix equation in Eq.~\ref{BetheSalpeterNoVeto}
we would like to write it in the form of an eigenvalue problem
yielding $t=1/u$ as a function of $\kappa^2$. This
is slightly complicated since the propagator segments involve
$\Sigma'$'s which appear together with factors of $\kappa^2$. By setting
$\chi=u\kappa^2$ we can manipulate the equation to isolate $t$ as the
eigenvalue, with solutions $t_n(\chi)$. This is achieved by rescaling
$\Psi_{\,\psfig{file=left.eps,height=.075in}\,}$,
$\Psi_{\,\psfig{file=right.eps,height=.075in}\,}$ by the denominator factor
common to all ${\cal D}$'s, yielding
\begin{eqnarray}
&&\alpha_l\left(1-\chi\Sigma_l'\right)
\left(1-\chi\Sigma_{M-l}'\right)
\Psi_{\,\psfig{file=right.eps,height=.075in}\,}'(l)
+\sum_{r=1}^{M-l-1}{\chi\over r^2}\alpha_{l+r}
\alpha^{2/(M-l-r)}\left(1-\chi\Sigma_{M-l-r}'\right)
\Psi_{\,\psfig{file=left.eps,height=.075in}\,}'(l+r)
\nonumber \\
&& \hskip 1cm
= t\left[\left(1-\chi^2\Sigma_l'\Sigma_{M-l}'\right)^2
\Psi_{\,\psfig{file=right.eps,height=.075in}\,}'(l)
-\sum_{r=1}^{M-l-2}{\chi\over r^2}\alpha_{l+r}
\left(1-\chi^2\Sigma_{l+r}'\Sigma_{M-l-r}'\right)
\Psi_{\,\psfig{file=right.eps,height=.075in}\,}'(l+r) \right.
\nonumber \\
&& \hskip 2cm
\left.+\sum_{r=1}^{M-l-1}{\chi^2\over r^2}\alpha^{2/(M-l-r)}
\left(1-\chi^2\Sigma_{l+r}'\Sigma_{M-l-r}'\right)
\Psi_{\,\psfig{file=left.eps,height=.075in}\,}'(l+r)\right]
\nonumber \\
&& \alpha_l\left(1-\chi\Sigma_l'\right)
\left(1-\chi\Sigma_{M-l}'\right)
\Psi_{\,\psfig{file=left.eps,height=.075in}\,}'(l)
+\sum_{r=1}^{l-1}{\chi\over r^2}\alpha_{l-r}
\alpha^{2/(l-r)}\left(1-\chi\Sigma_{l-r}'\right)
\Psi_{\,\psfig{file=right.eps,height=.075in}\,}'(l-r)
\nonumber \\
&& \hskip 1cm
= t\left[\left(1-\chi^2\Sigma_l'\Sigma_{M-l}'\right)^2
\Psi_{\,\psfig{file=left.eps,height=.075in}\,}'(l)
-\sum_{r=1}^{l-2}{\chi\over r^2}\alpha_{l-r}
\left(1-\chi^2\Sigma_{l-r}'\Sigma_{M-l+r}'\right)
\Psi_{\,\psfig{file=left.eps,height=.075in}\,}'(l-r)\right.
\nonumber \\
&& \hskip 2cm
\left.+\sum_{r=1}^{l-1}{\chi^2\over r^2}\alpha^{2/(l-r)}
\left(1-\chi^2\Sigma_{l-r}'\Sigma_{M-l+r}'\right)
\Psi_{\,\psfig{file=right.eps,height=.075in}\,}'(l-r)\right], 
\end{eqnarray} 
where $\alpha_l\equiv\alpha^{M/l(M-l)}$.

\begin{figure}[ht]
\centerline{
$
\begin{array}[c]{c}
\psfig{file=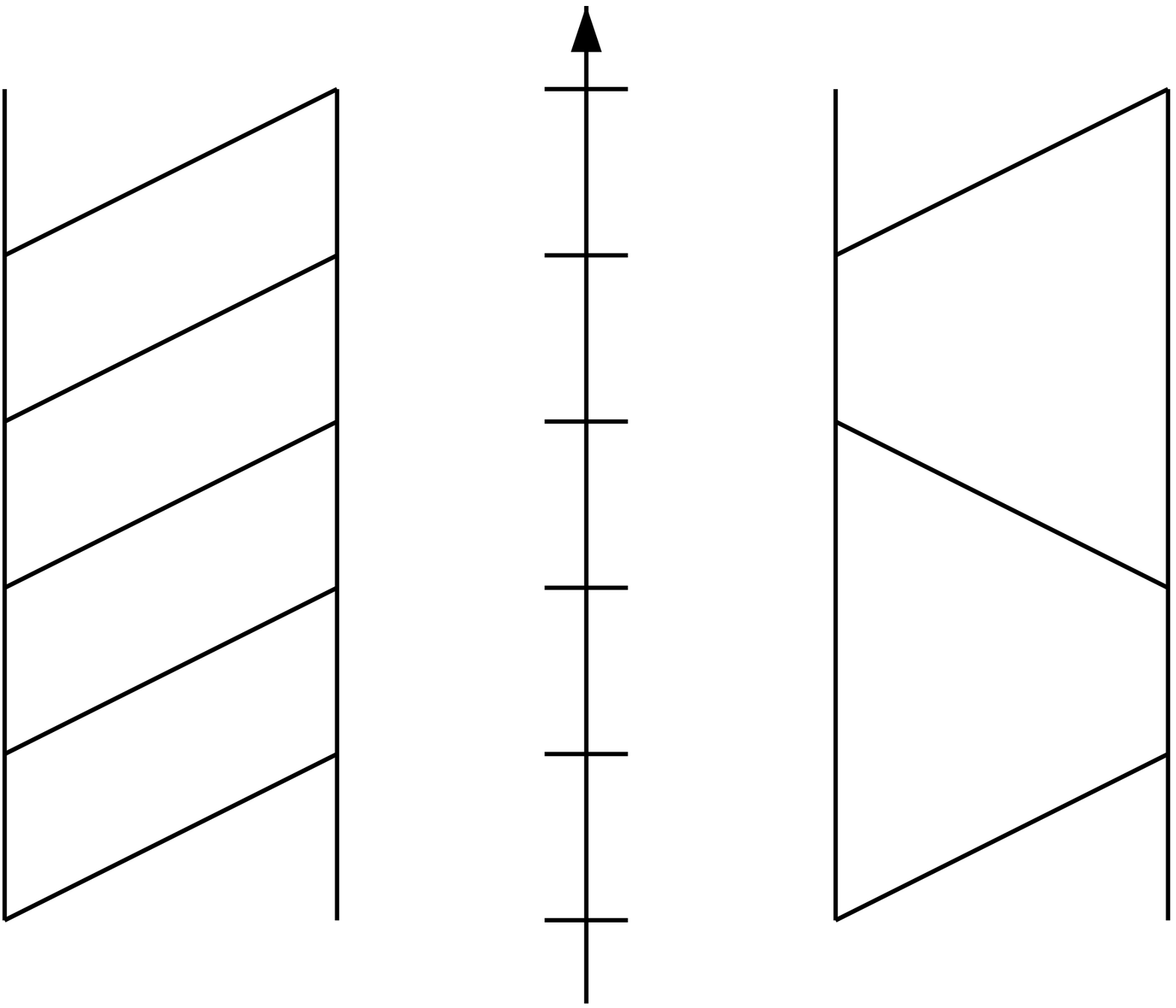,height=2.0in}
\end{array}
\quad\Longrightarrow\quad
\begin{array}[c]{c}
\psfig{file=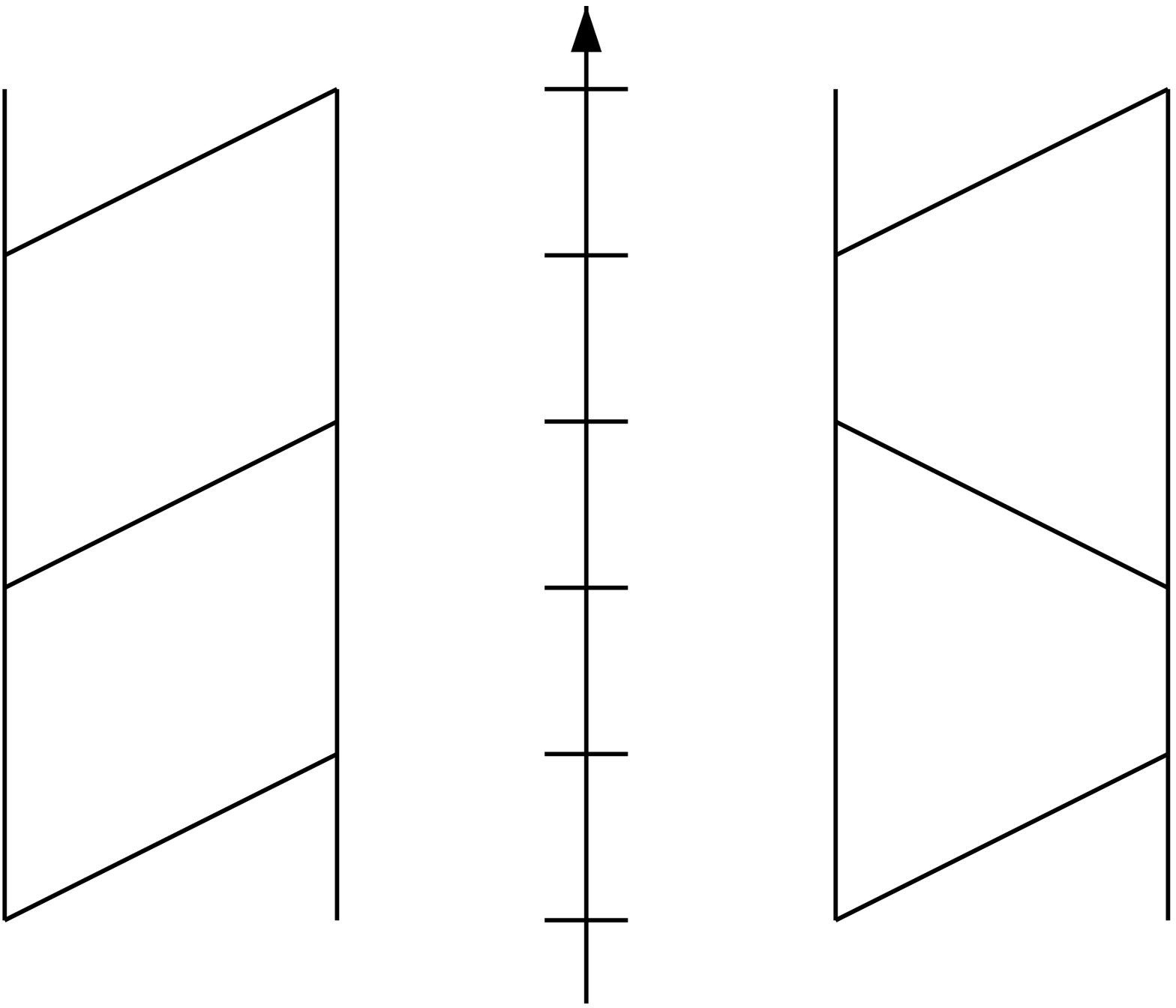,height=2.0in}
\end{array}
$
}
\caption[]{Asymmetry in the densest configuration of
exchanges in the same sense and opposite sense. The double arrow points to
the effect of implementing the veto.}
\label{asymrungs}
\end{figure}

\begin{figure}[ht]
\centerline{\psfig{file=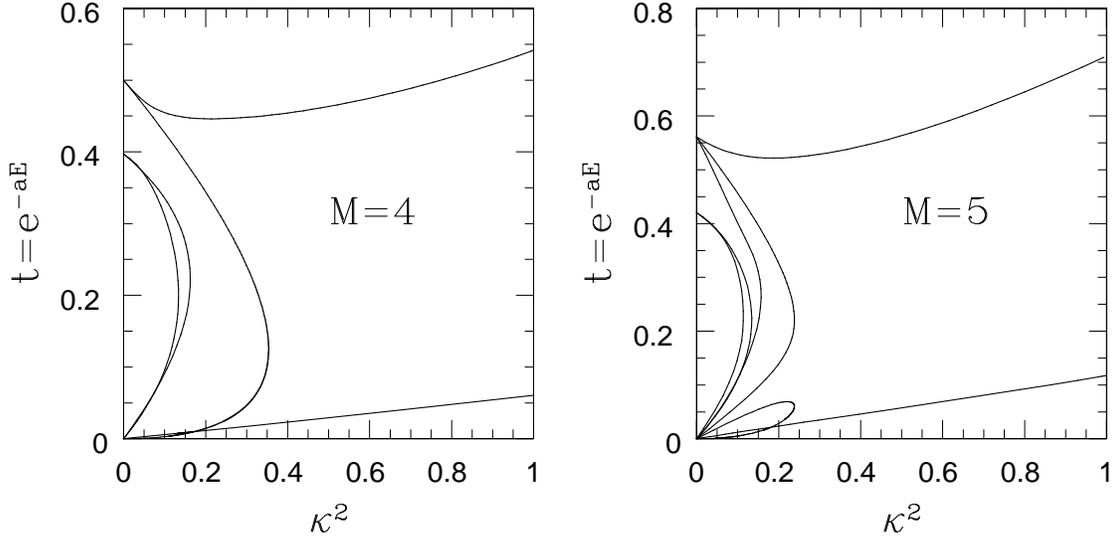,width=6.5in}}
\caption[]{Real eigenvalues using the Bethe-Salpeter method for $M=4,5$.
All solutions of the single time step method, see Fig.~\ref{m45sts}, are
reproduced, but additional spurious solutions are present.}
\label{m45noveto}
\end{figure}

This discretized equation has roughly twice the complexity of a
straightforward discretization of the 't Hooft equation. The reason
is that a rung propagating forward from left to right
can couple to subsequent evolutions forbidden to a
rung from right to left (and vice versa). 
See Fig.~\ref{asymrungs}, for the graphs responsible
for this asymmetry. 
This is the reason we had to introduce
a two-component Bethe-Salpeter wave function. An immediate 
consequence is that at $\kappa=0$ each energy value is at
least doubly degenerate, including the ground state. This
feature is evident in Fig.~\ref{m45noveto} where the solutions of the
BS equation are displayed for $M=4,5$. 
All of the solutions seen in
Fig.~\ref{m45sts} are present, but in addition there are
extra spurious solutions. For example, with $M=4$, there
is a second
curve emerging from the $\kappa=0$ ground state eigenvalue. For $\kappa>0$
this extra eigenvalue curve lies below (in $t$) and well separated
from the true ground level curve for all coupling. Similarly,
for other values of $M$ the Bethe-Salpeter method consistently
reproduces all the solutions of the transfer matrix method, but
it also adds spurious solutions due to the two-component
nature of the wave function.

\begin{figure}[ht]
\centerline{\psfig{file=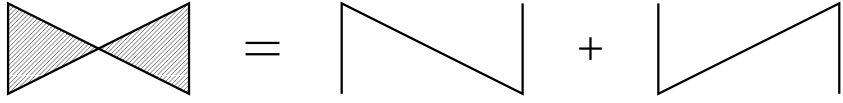,height=.5in}}
\caption[]{With successive emissions and absorptions vetoed, the
two types of exchanges can be combined in a single rung.}
\label{symrungs}
\end{figure}

One way to avoid these unwanted solutions is to slightly modify
the discretized Feynman rules so that the rung will attach
to the same lines whichever  way the exchanged
gluon propagates. As seen in Fig.~\ref{asymrungs}, the asymmetry
stems from the possibility of consecutive gluon
emissions (absorptions) on immediately successive time steps. 
If this
possibility is disallowed, the basic exchange rung can
be taken to be the sum of the two different exchanges as in 
Fig.~\ref{symrungs}. 
In addition to removing unwanted solutions
this veto rule also leads to simpler equations, with a more
transparent continuum limit. As we shall see
in the next section, it also produces a more physical
strong coupling behavior than our original discretization.  

\setcounter{equation}0
\renewcommand{\theequation}{\thesection.\arabic{equation}}
\section{Bethe-Salpeter with Veto}
\label{sec5}

The Bethe-Salpeter equation for the discretized 't Hooft model, with
the veto imposed as described at the end of the previous section, is
\begin{eqnarray}
\Psi(l)&=&\sum_{r=1}^{M-l-1} {\kappa^2\over r^2}u\alpha^{1/l+1/(M-l-r)}
{\cal D}_{q\bar{q}}(l+r)\Psi(l+r)
\nonumber \\
&&\qquad\qquad
+\sum_{r=1}^{l-1} {\kappa^2\over r^2}u\alpha^{1/(l-r)+1/(M-l)}
{\cal D}_{q\bar{q}}(l-r)\Psi(l-r),
\end{eqnarray}
where ${\cal D}_{q\bar{q}}$ is defined in Eq.~\ref{MesonProp}. After
re-indexing both sums the equation can be written as
\begin{eqnarray}
\Psi(l)=\sum_{r=l+1}^{M-1} {u\kappa^2\over (l-r)^2}\alpha^{1/l+1/(M-r)}
{\cal D}_{q\bar{q}}(r)\Psi(r)
+\sum_{r=1}^{l-1} {u\kappa^2\over (l-r)^2}\alpha^{1/r+1/(M-l)}
{\cal D}_{q\bar{q}}(r)\Psi(r).
\label{thoofteqveto}
\end{eqnarray}
By imposing the veto we have reduced the rank of the eigenvalue problem from
$2(M-2)$ to $M-1$. 

The new discretized equation is much easier to analyze in the
formal continuum limit $M\to\infty$ than the original.  
First define $\Phi(r)\equiv {\cal D}_{q\bar q}(r)\Psi(r)$,
and rearrange Eq.~\ref{thoofteqveto} to read
\begin{eqnarray}
K(l)\Phi(l)&\equiv&\left({1\over u{\cal D}_{q\bar{q}}(l)}-
\kappa^2(\Sigma^\prime_l+\Sigma^\prime_{M-l})\right)\Phi(l)
\label{thoofteqveto2}\\
&=&\sum_{r=l+1}^{M-1} 
{\kappa^2\over (l-r)^2}\alpha^{1/l+1/(M-r)}
(\Phi(r)-\Phi(l))
+\sum_{r=1}^{l-1} {\kappa^2\over (l-r)^2}\alpha^{1/r+1/(M-l)}(\Phi(r)-\Phi(l)).
\nonumber
\end{eqnarray}
To formally
examine the continuum limit we suppose that each discrete $p^+$
variable is large putting each $l\to xM$,\footnote{Of course even for
$M$ large the equation does contain terms where $l$ and $M-l$ are
small (i.e. close to 1). In order for these contributions to not
affect the solution to the continuum Bethe-Salpeter equation, the
wavefunction must vanish at the endpoints. We shall see how this occurs
when we evaluate the numerics later.} 
and take $M\to\infty$ at fixed $x$. Then the right hand side
of Eq.~\ref{thoofteqveto2} is set up to go to $1/M$ times the r.h.s. of
the continuum 't Hooft equation:
\begin{eqnarray}
&&\hskip-2cm\sum_{r=l+1}^{M-1} {\kappa^2\over (l-r)^2}\alpha^{1/l+1/(M-r)}(\Phi(r)-\Phi(l))
+\sum_{r=1}^{l-1} {\kappa^2\over (l-r)^2}\alpha^{1/r+1/(M-l)}(\Phi(r)-\Phi(l))
\nonumber\\
&&\hskip2cm\to{\kappa^2\over M}P\int_0^1 dy {\Phi(y)-\Phi(x)\over (y-x)^2}.
\label{thoofteqrhs}
\end{eqnarray}
Clearly, $u$ must be chosen so that the l.h.s. is also of order $1/M$.
Next, it is easy to verify that $\Sigma^\prime_l=\alpha^{1/l +1/(M-l)}
(\pi^2/6-1/l+{\cal O}(\ln l/l^2))$, so that the inverse 
propagator can be simplified, neglecting terms of order $\ln M/M^2$,
\begin{eqnarray}
{1\over u_l{\cal D}_{q\bar q}(l)}\sim {1\over u_l^2}
-\kappa^4\left({\pi^2\over6}
-{M\over2l(M-l)}\right)^2-{1\over u_l}
{1-u_l\kappa^2(\pi^2/6-M/2l(M-l))\over1+u_l\kappa^2(\pi^2/6-M/2l(M-l))},
\end{eqnarray}
where we have defined $u_l=u\alpha^{M/l(M-l)}$.
The factor $K$ multiplying $\Phi$ on the l.h.s. of Eq.~\ref{thoofteqveto2} 
can now be simplified to
\begin{eqnarray}
K(l)&\sim&\alpha^{M/l(M-l)}\left[{1\over u_l^2}-{\kappa^4\pi^4\over36}
-{1\over u_l}
{1-u_l\kappa^2(\pi^2/6-M/2l(M-l))\over1+u_l\kappa^2(\pi^2/6-M/2l(M-l))}
-{\kappa^2\pi^2\over3}\right.\nonumber\\
&&\hskip2.75in\mbox{}\left.
+{\kappa^4\pi^2\over6}{M\over l(M-l)}+\kappa^2{M\over l(M-l)}\right]
\nonumber\\
&\sim&\alpha^{M/l(M-l)}\left[{1\over u_l^2}-{\kappa^4\pi^4\over36}
-{1\over u_l}
{1-u_l\kappa^2\pi^2/6\over 1+u_l\kappa^2\pi^2/6}
-{\kappa^2\pi^2\over3}\right.\nonumber\\
&&\mbox{}\hskip2in\left.+{M\over l(M-l)}\left[{\kappa^4\pi^2\over6}+\kappa^2
-{\kappa^2\over(1+u_l\kappa^2\pi^2/6)^2}\right]\right]
\end{eqnarray}
Now write $u=u_0e^{a\Delta}$, where $a\Delta$ will be determined
to be of order $1/M$, so that  
$u_l=u_0(1+a\Delta+(M/l(M-l))\ln\alpha)$ to order $1/M$. 
Then $u_0$ must satisfy
\begin{eqnarray}
f(u_0)\equiv{1\over u_0^2}-{\kappa^4\pi^4\over36}-{1\over u_0}
{1-u_0\kappa^2\pi^2/6\over1+u_0\kappa^2\pi^2/6}-{\kappa^2\pi^2\over3}=0.
\label{u0}
\end{eqnarray} 
Then the continuum limit reads
\begin{eqnarray}
\left[a\Delta+{1\over Mx(1-x)}\left\{
\ln\alpha+{1\over u_0f^\prime(u_0)}\left[{\kappa^4\pi^2\over6}+\kappa^2
-{\kappa^2\over(1+u_0\kappa^2\pi^2/6)^2}\right]\right\}\right]\Phi(x)
&& \label{thoofteqcont} \\
&&\hskip-1.5in=\quad{\kappa^2\over Mu_0f^\prime(u_0)}P\int_0^1 dy {\Phi(y)-\Phi(x)\over (y-x)^2}.\hskip0in\nonumber
\end{eqnarray}
The energy of the system is $E=(\ln u_0)/a+\Delta$, but the
divergent first term is simply a physically irrelevant
$M$ independent constant, so it is consistent to identify 
$P^-=\Delta$. Then ${\cal M}^2=2P^+P^-=2Mm\Delta=2MT_0a\Delta$.
We also identify 
$$\mu^2=-2T_0\left\{
\ln\alpha+{1\over u_0f^\prime(u_0)}\left[{\kappa^4\pi^2\over6}+\kappa^2
-{\kappa^2\over(1+u_0\kappa^2\pi^2/6)^2}\right]\right\},$$ 
and we obtain the continuum 't~Hooft equation
\begin{eqnarray}
\left[{\cal M}^2-\mu^2{1\over x(1-x)}\right]\Phi(x)
&=&{2T_0\kappa^2\over u_0f^\prime(u_0)}P\int_0^1 dy {\Phi(y)-\Phi(x)\over (y-x)^2}\nonumber\\
&=&{g_s^2N_c\over2\pi u_0f^\prime(u_0)}P\int_0^1 dy {\Phi(y)-\Phi(x)\over (y-x)^2}.
\label{thoofteqcont2}
\end{eqnarray}
Comparing with Eq.~\ref{thoofteq}, we see that the only effect
on the continuum limit of keeping $T_0$ finite is a finite
renormalization of the gauge coupling $g^2\to-g^2/u_0f^\prime(u_0)$,
and a coupling constant dependent shift in $\mu^2$.
Thus, the only requirement for identical continuum physics is that
$u_0f^\prime(u_0)$ be negative. Since $\alpha$ is a free parameter,
we can access all positive values of $\mu^2$ by tuning it.

Eq.~\ref{u0} implicitly relates $u_0$ to $\kappa$ via a cubic equation.
Instead of solving this equation, it is more illuminating to use
it to relate $u_0$ to the combination $\eta\equiv u_0\kappa^2\pi^2/6$
\begin{eqnarray}
u_0={(1-\eta^2)(1+\eta)\over1+\eta+2\eta^2},\qquad
\kappa^2={6\eta\over u_0\pi^2}={6\eta(1+\eta+2\eta^2)
\over(1-\eta^2)(1+\eta)\pi^2}.
\label{ukappaeta}
\end{eqnarray}
We can also obtain the charge renormalization factor $u_0f^\prime(u_0)$
in terms of $\eta$:
\begin{eqnarray}
u_0f^\prime(u_0)=-{(1+\eta+2\eta^2)(1+\eta+7\eta^2-\eta^3)
\over(1-\eta^2)^2(1+\eta)^2},
\end{eqnarray}
the effective coupling in the 't~Hooft equation
\begin{eqnarray}
{g_{\rm eff}^2N_c\over\pi}=-{2\kappa^2T_0\over u_0f^\prime(u_0)}={12\eta(1-\eta^2)(1+\eta)T_0\over\pi^2
(1+\eta+7\eta^2-\eta^3)},
\label{effcoupling}
\end{eqnarray}
and the renormalized mass parameter 
\begin{eqnarray}
\mu^2=\mu_0^2+{12\eta^2(3+\eta^2)T_0\over\pi^2
(1+\eta+7\eta^2-\eta^3)},
\label{effmass}
\end{eqnarray}
where we have used  $\alpha=e^{-\mu_0^2/2T_0}$. 

As a check, note that the continuous time limit corresponds to 
$T_0\to\infty$ or $\kappa^2\to0$, whence $u_0\to1$ and $\eta\to0$.
Then the effective coupling Eq.~\ref{effcoupling} goes to 
$12T_0\eta/\pi^2=2T_0\kappa^2=g^2N_c/\pi=g_s^2N_c/2\pi$
as it should. Next, with discrete time, we see that, in
order to have real energy and $\kappa$ ($u_0>0$ and $\kappa^2>0$),
we must place the restriction $0<\eta<1$. Small $\kappa$
corresponds to small $\eta$, and large $\kappa$ corresponds
to $\eta$ near unity. Interestingly, we note that the
effective coupling in the 't~Hooft equation is small in {\it both}
the small and large $\kappa$ regimes.

It is easy to understand the small effective coupling at 
large $\kappa$ in terms of our discrete time Feynman
diagrams. With discrete time, $\kappa^2\to\infty$ 
causes the diagrams with a maximal number of
powers of $\kappa^2$ per time step to dominate. For example
the $q\bar q$ propagator ${\cal D}_{q\bar q}$  behaves in this limit as
\begin{eqnarray}
{\cal D}_{q\bar{q}}(l)
\sim{u\alpha^{M/l(M-l)}
\over 1-u^2\kappa^4\Sigma_l'\Sigma_{M-l}'}=
u\alpha^{M/l(M-l)}
\sum_{k=1}^\infty (u\kappa^2)^{2k}(\Sigma_l'\Sigma_{M-l}')^k,
\label{strongMesonProp}
\end{eqnarray}
so that the propagator for $2k+1$ time steps is $\alpha^{M/l(M-l)}
(\kappa^4\Sigma_l'\Sigma_{M-l}')^k\to(\kappa^2\pi^2/6)^{2k}$
in the continuum limit. We see that away from
the endpoints there is a factor of $\kappa^2\pi^2/6$ per time step
in the continuum limit,
which corresponds to each quark propagating exactly one
time unit between interactions. Since this is the eigenvalue
of the transfer matrix, we immediately infer the strong coupling
value of $u_0=6/\kappa^2\pi^2$. Because of our veto,
every exchange between quark lines occupies precisely two time steps
and contributes only a single factor of $\kappa^2$. Thus each
exchange costs a relative factor of $1/\kappa^2$ in the
strong coupling limit, and this relative factor is proportional to
the effective coupling in the 't~Hooft equation. More precisely,
separating out the factor corresponding to the strong
coupling propagation of the quark and anti-quark for two time steps,
we have $\kappa^2=(\kappa^2\pi^2/6)^2(36/\kappa^2\pi^4)$,
so the effective coupling for a single exchange is
$36/\kappa^2\pi^4$ for large $\kappa$, in accord with the $\eta\to1$ limit
of Eqs.~\ref{ukappaeta},~\ref{effcoupling}.

Now we turn to a numerical analysis of our
discretized dynamics in order to understand 
how the continuum limit is approached in practice.
As with the no-veto case in section~\ref{sec4} we
can write this equation as an eigenvalue problem by rescaling
$\Psi$ and isolating the eigenvalue $t$ as a function of $\chi\equiv
u\kappa^2$. The resulting eigenvalue problem to solve is
\begin{eqnarray}
t\Phi(l) &=&
{\alpha_l\over\left(1-\chi^2\Sigma_l'\Sigma_{M-l}'\right)}
\left[
{\left(1-\chi\Sigma_l'\right)
\left(1-\chi\Sigma_{M-l}'\right)\over
\left(1-\chi^2\Sigma_l'\Sigma_{M-l}'\right)}\Phi(l)
+\chi\sum_{r=1}^{l-1} {\alpha^{1/r+1/(M-l)}\over (l-r)^2}\Phi(r)
\right. \nonumber \\
&&\hskip 2.5in \left.
+\chi\sum_{r=l+1}^{M-1} {\alpha^{1/l+1/(M-r)}\over (l-r)^2}
\Phi(r)\right].
\label{BSveto}
\end{eqnarray}
We use numerical procedures in MAPLE and MATLAB to find the
eigenvalues $t_n(\chi)$ of the matrix on the right hand side of this
equation as a function of $\chi$. The value of $\kappa^2$ is different
for each $t_n$ since $\kappa^2=\chi t_n$. However by varying
$0\leq\chi\leq\infty$ we can generate the real solutions, $t_n$, for 
all $\kappa^2$. In order to solve for complex $t_n$'s we
would need to vary $\chi$ in the complex plane rather than just
over positive real numbers.

\begin{figure}[htp]
\centerline{\psfig{file=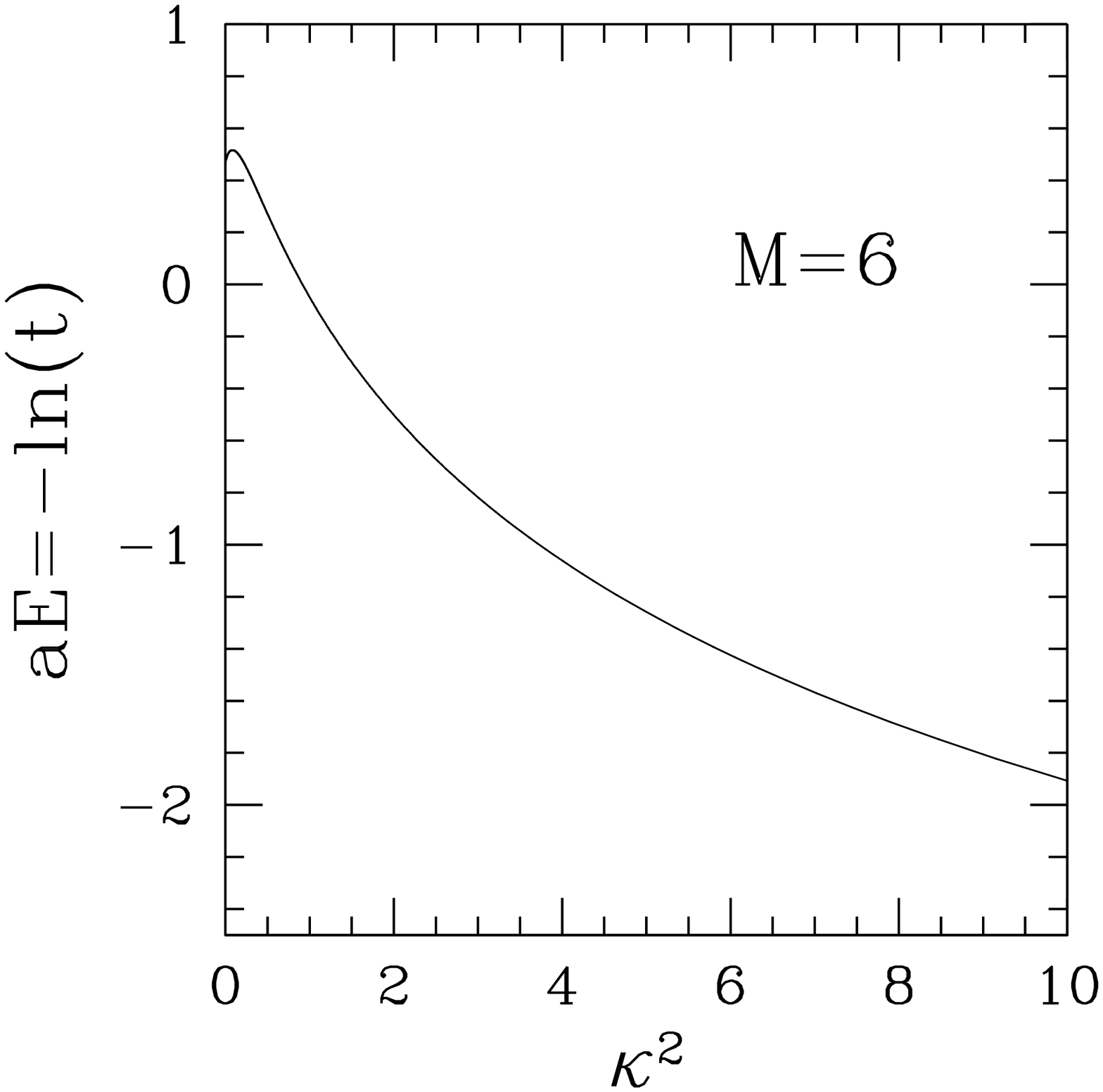,height=2.9in}\qquad
\psfig{file=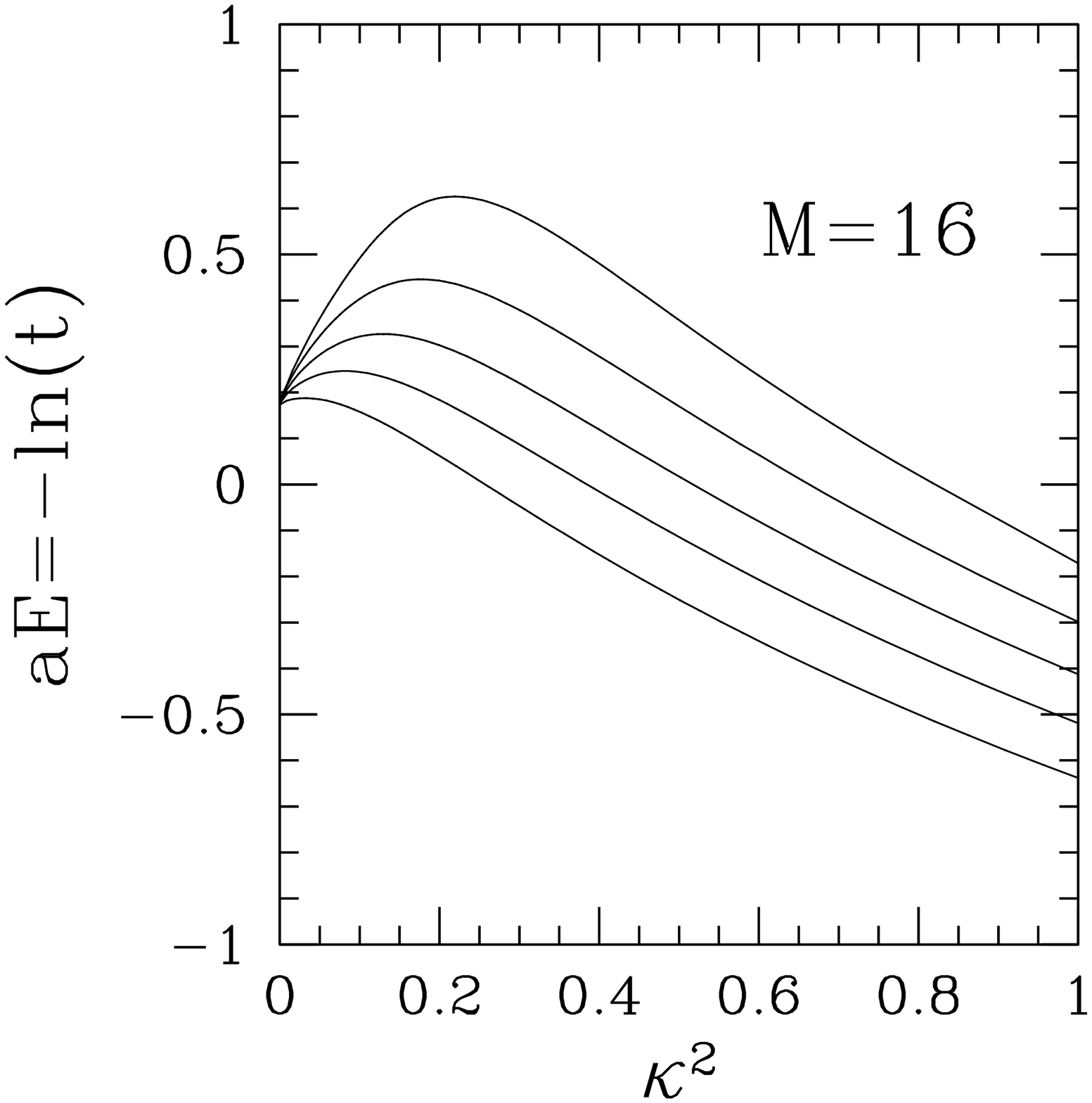,height=2.9in}}
\centerline{\psfig{file=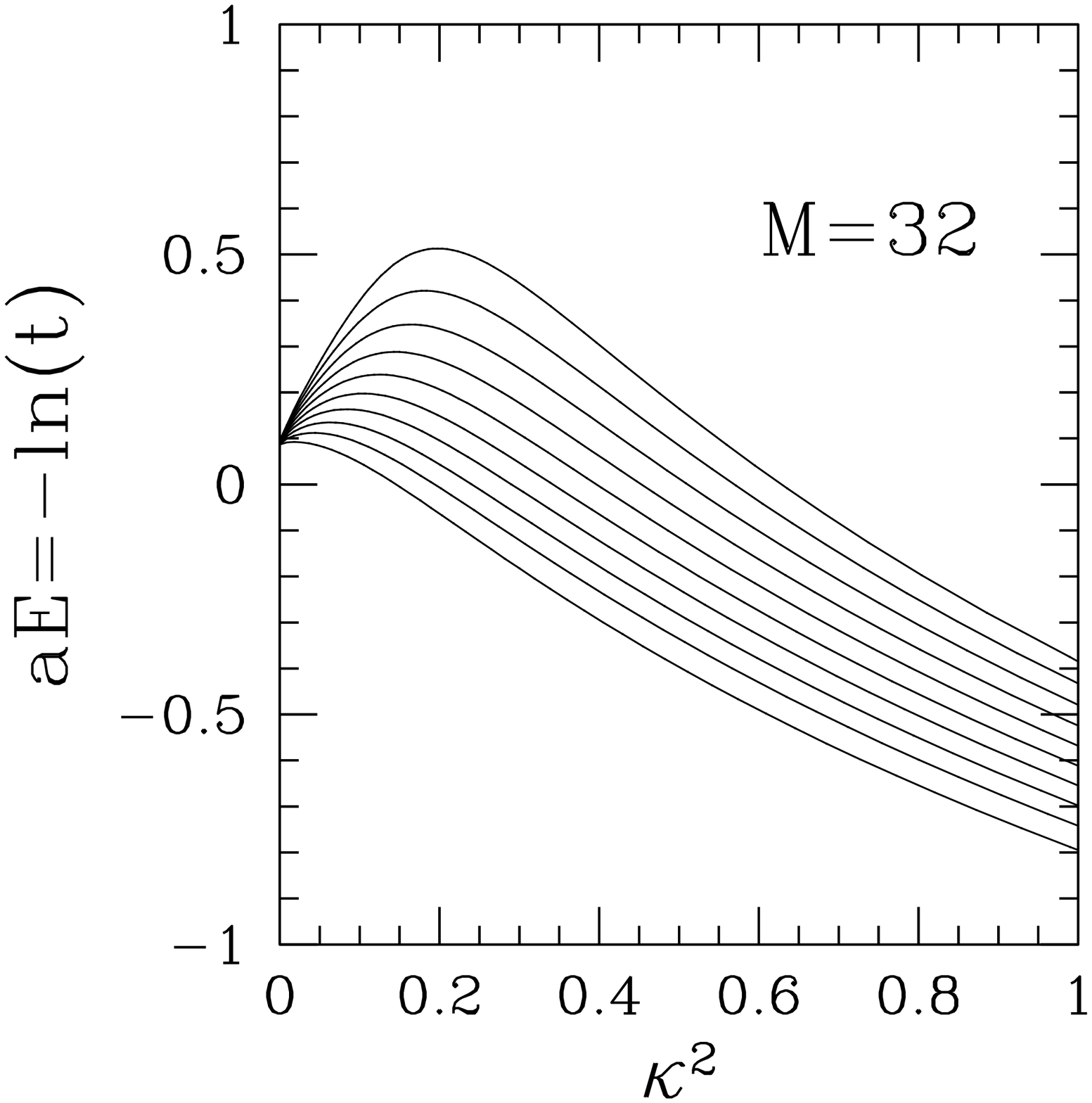,height=2.9in}\qquad
\psfig{file=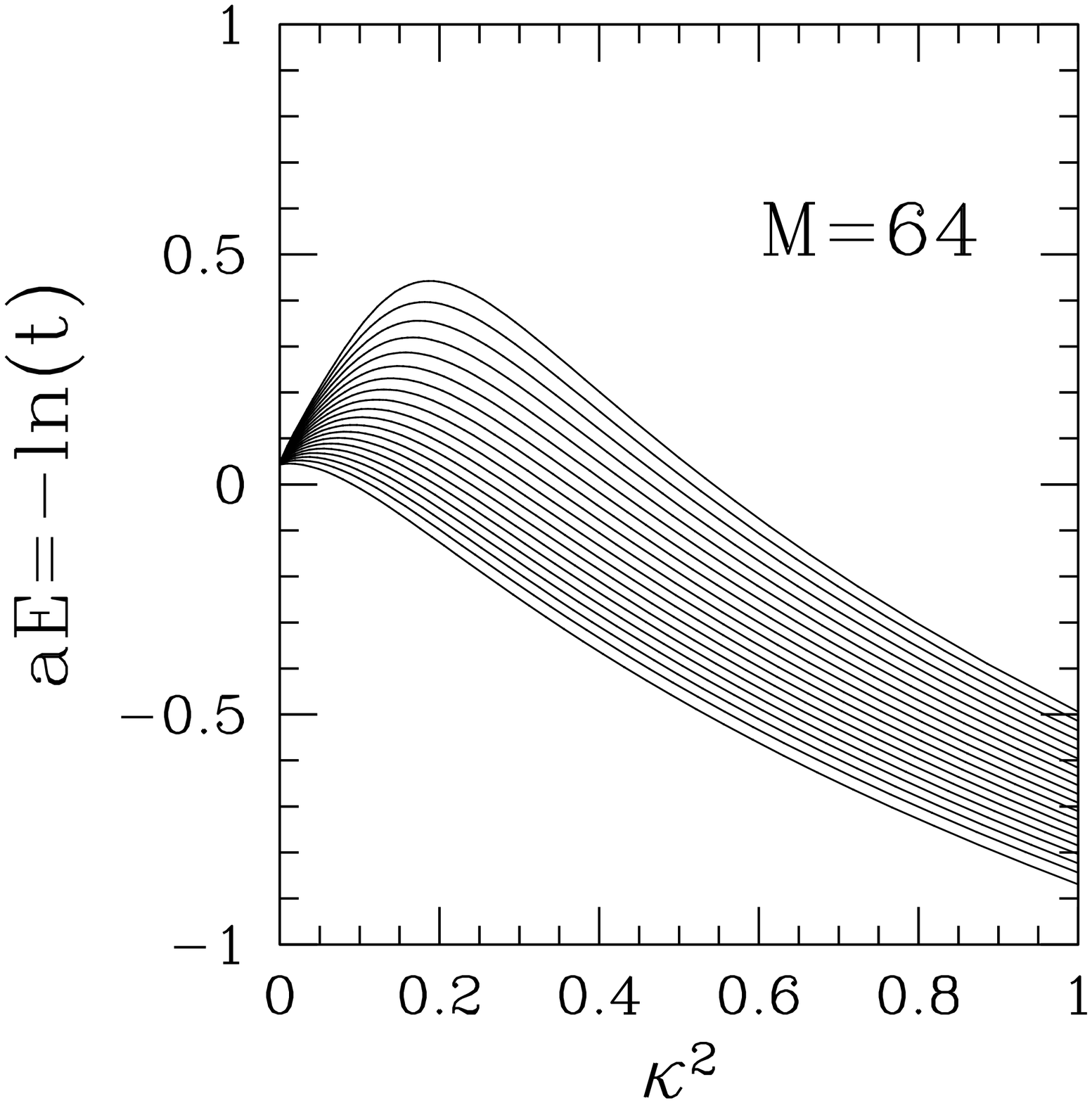,height=2.9in}}
\caption[]{Plots of the lowest lying energy eigenstates
of the Bethe-Salpeter equation with the veto for $M=6,16,32,64$. Other
states which occur at higher energies than those displayed have been
omitted.} 
\label{vetographs}
\end{figure}
The problem of contamination of the lowest lying states by complex
solutions has been solved by our veto prescription: 
The lowest lying state for $M=6$ for Eq.~\ref{BSveto} remains
intact for all coupling $\kappa^2$, see Fig.~\ref{vetographs},
which should be compared against Fig.~\ref{M56singletimestep} 
where the lowest lying state was only the ground state for 
$\kappa^2\lesssim3$. When we analyze Eq.~\ref{BSveto} for increasing $M$
(see Fig.~\ref{vetographs} for $M=16,32,64$) we see that the number of
low lying states that remain uncrossed for all couplings increases
with increasing $M$. We also see that the spacing between these states
decreases as $M$ increases. Recall that the solutions in 
Fig.~\ref{vetographs} have been generated for $\alpha=0.5$.

In order to compare our numerical results for large values of $M$
(hopefully close to the continuum limit) with the numerical results of
't Hooft~\cite{thooftmodel} we solve the Bethe-Salpeter equation in
Eq.~\ref{BSveto} for $\kappa^2=0.5$ and 
\begin{eqnarray}
\alpha=1.16433 \qquad \alpha=1.04167 \qquad \alpha=0.70930.
\end{eqnarray}
These three choices of $\alpha$ correspond to values of 
't Hooft parameter, $\tilde{\mu}^2\equiv\pi\mu^2/g_{\rm eff}^2N_c$, 
taken to be $0$, $1$
and $2.11^2$ respectively. These values of $\tilde{\mu}$ 
were used in~\cite{thooftmodel}. Fixing $\kappa^2$ is
equivalent to fixing $\eta$, $u_0$ and $g_{\rm eff}^2N_c/\pi$, thus
choosing a value for $\tilde{\mu}$ determines $\alpha$ in Eq.~\ref{effmass}.

\begin{figure}[htp]
\centerline{\psfig{file=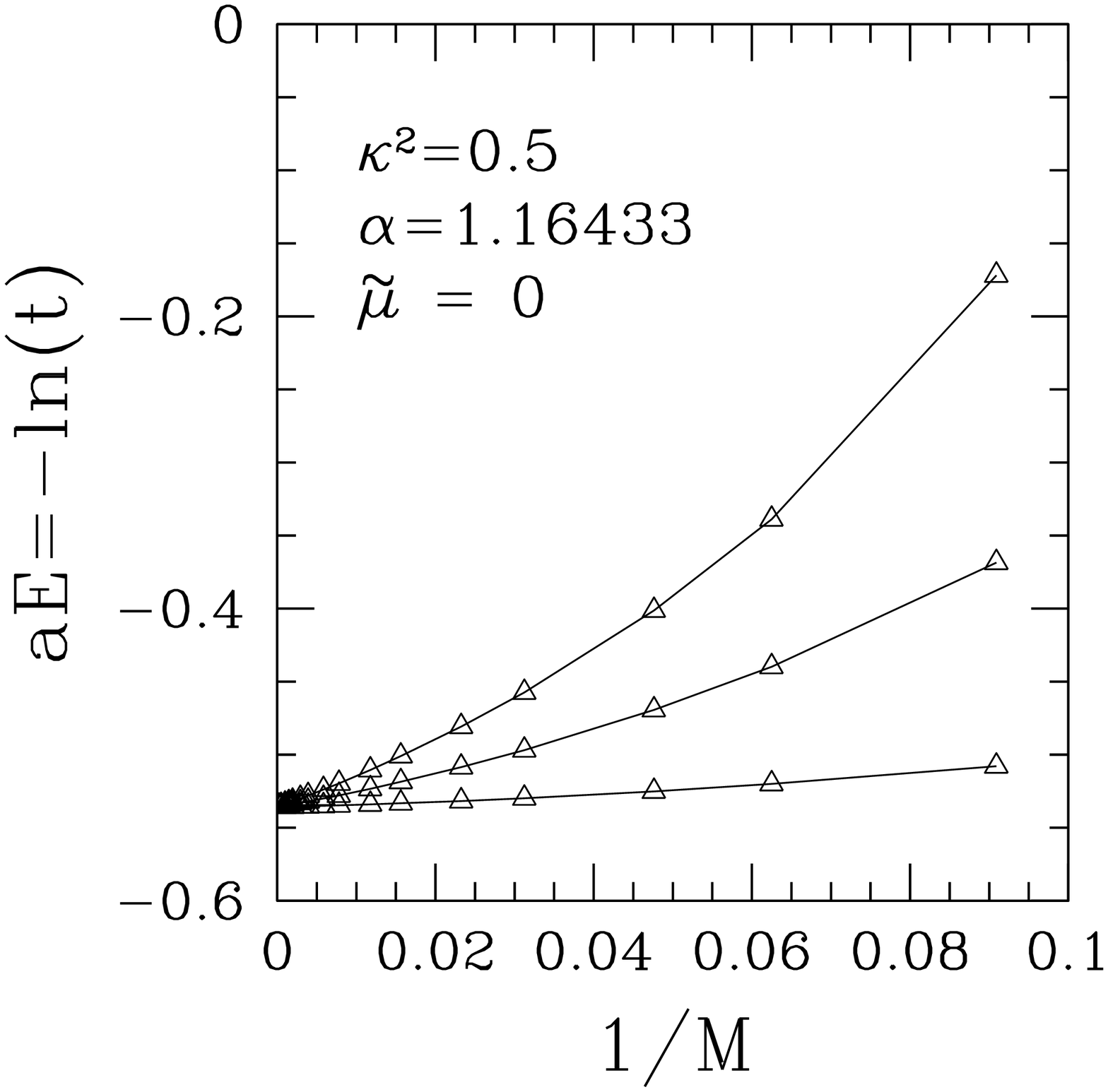,height=2.9in}\qquad
\psfig{file=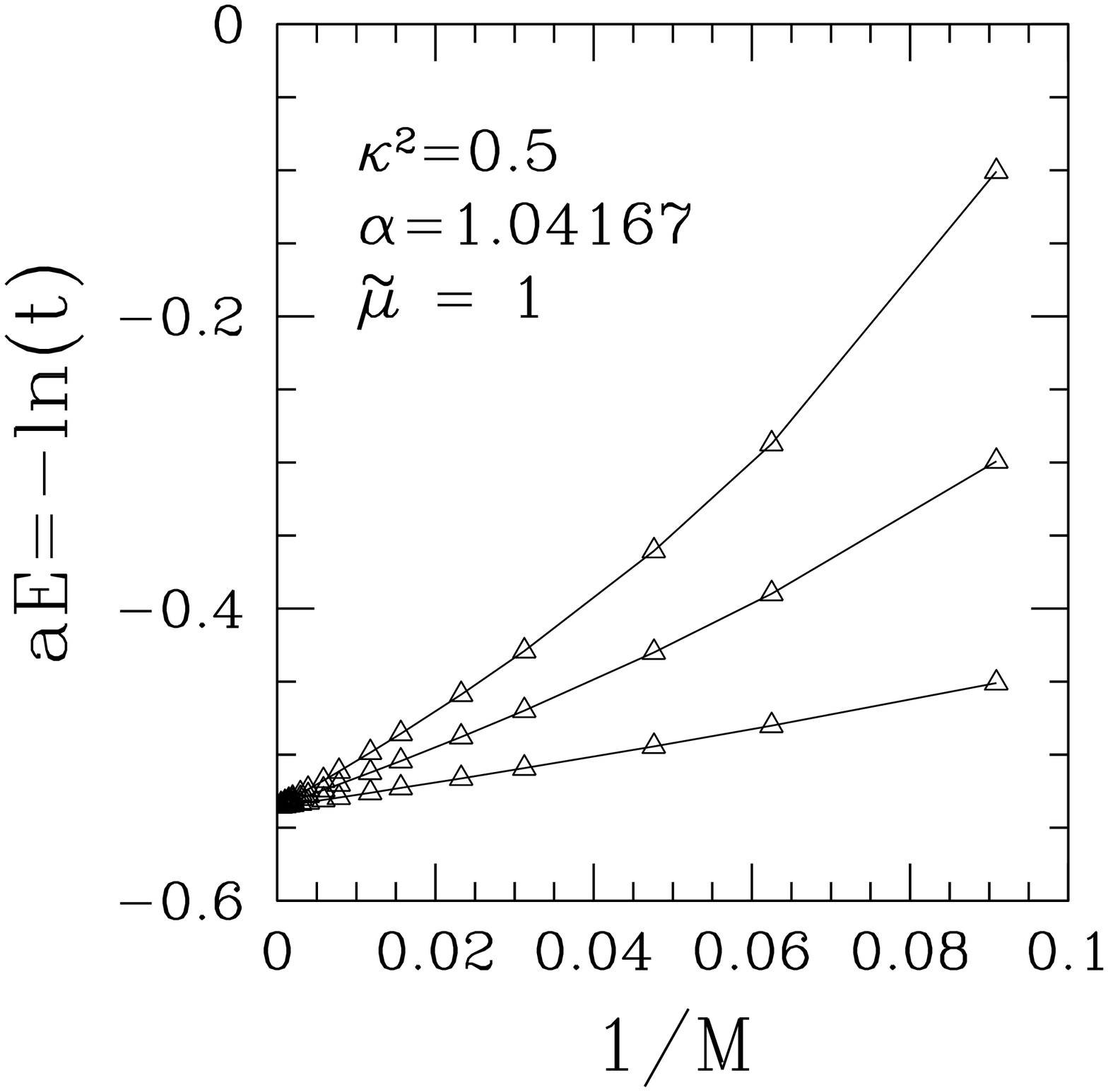,height=2.9in}}
\centerline{\psfig{file=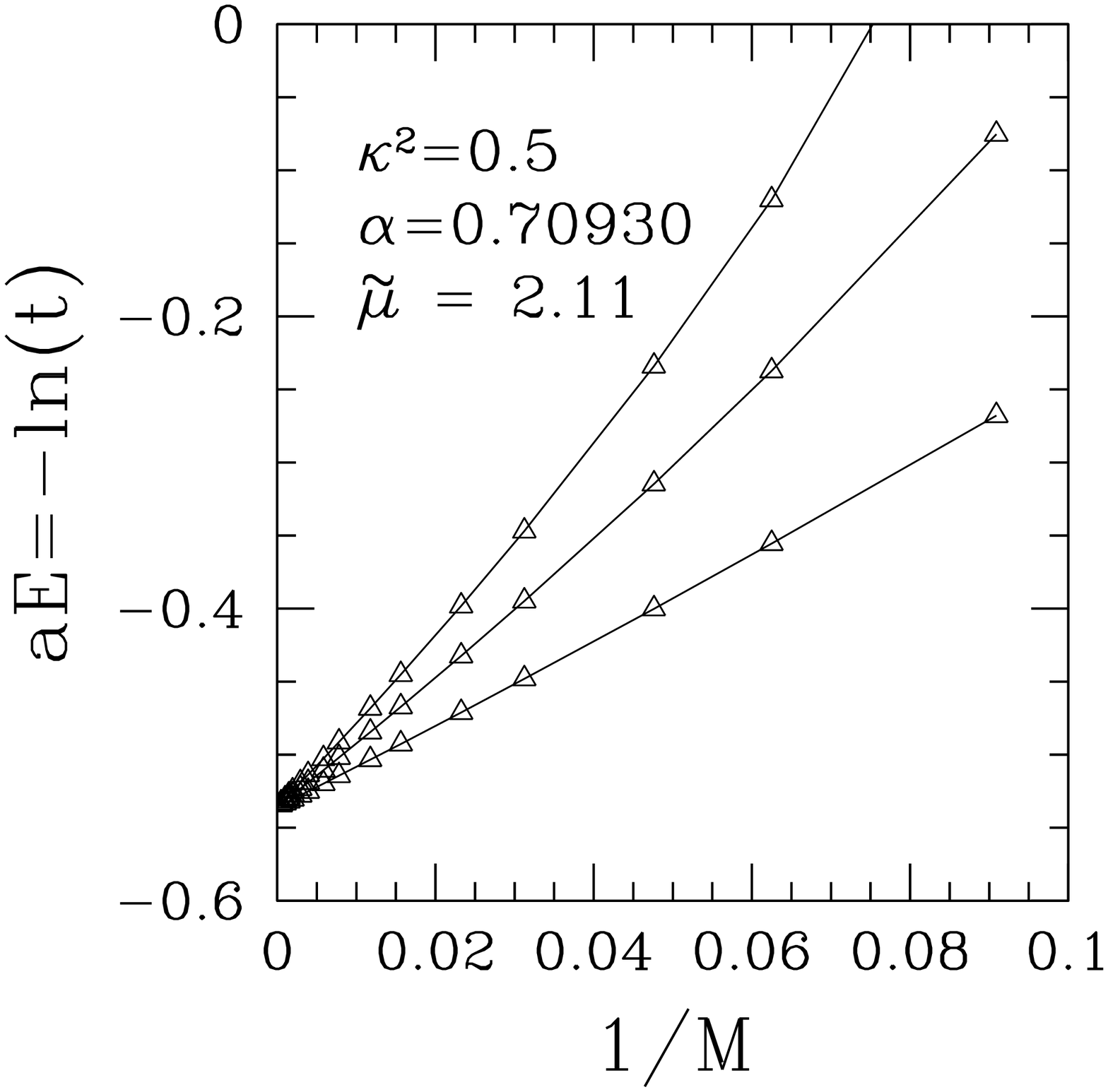,height=2.9in}}
\caption[]{Plots of the three lowest lying states against $1/M$. The
three graphs correspond to choices of $\alpha$ and $\kappa^2$ such
that the continuum limit of $\tilde{\mu}^2\equiv
\pi\mu^2/g_{\rm eff}^2N_c=0,1,(2.11)^2$
so that we can compare these results against those of 
't Hooft~\cite{thooftmodel}.}
\label{1overMcurves}
\end{figure}
As we can see in Fig.~\ref{1overMcurves} plots of the three lowest
lying energy levels against $1/M$ show curves that become linear with
increasing $M$. These results can be fitted to the functional form
\begin{eqnarray}
aE = \ln(u_0) + {c_1\over M} \exp\left({c_2\over M}+{c_3\over M^2}\right),
\label{FittedEquation}
\end{eqnarray}
where $c_2$ and $c_3$ parameterize the departure from $1/M$ behaviour
away from large $M$. We used the data of Fig.~\ref{1overMcurves} in
the range $128\leq M\leq 2048$ to fit this equation. With the fitted
value of $c_1$ we can calculate the mass square of the corresponding
bound state. As discussed previously, the $M$ independent term in
Eq.~\ref{FittedEquation} is dropped in identifying $P^-$. Since 
\begin{eqnarray}
{\cal M}^2 = 2P^+P^- = 2MT_0 (aE-\ln{u_0}),
\end{eqnarray}
we have, for $\kappa^2=0.5$, 
\begin{eqnarray}
{\cal M}^2 = {2c_1\over0.22265} + \ldots,
\label{MassSquare}
\end{eqnarray}
in units of $g_{\rm eff}^2N_c/\pi$. The results of the fits are
tabulated in Table~\ref{fits} against the results of 't
Hooft~\cite{thooftmodel}. 
\begin{table}[ht]
\begin{center}
\begin{tabular}{|c||c|c||c|c||c|c|}
\hline
& $\tilde{\mu}=0$ & 't Hooft & 
$\tilde{\mu}=1$ & 't Hooft &
$\tilde{\mu}=2.11$ & 't Hooft \\
\hline
ground state &  0.72 &    0 &  7.25 &  7.2 & 24.23 & 24.1 \\
1st          &  7.57 &  5.9 & 17.26 & 17.3 & 38.17 & 38.1 \\
2nd          & 16.21 & 14.3 & 27.06 & 27.2 & 49.98 & 49.8 \\
\hline
\end{tabular}
\end{center}
\caption[]{Comparison of numerical fits for $128\leq M\leq 2048$ (for 
$\tilde{\mu}=0$ we used $128\leq M\leq 4096$)
in order to determine the bound state mass squared in units of 
$g_{\rm eff}^2N_c/\pi$ for our discretized theory compared against the
numerical results of the conventional continuous time approach of 't
Hooft.}
\label{fits}
\end{table}
We see that for $\tilde{\mu}=1$ and $2.11$, the results of our 
discretization match
quite well those of~\cite{thooftmodel}. However, for
$\tilde{\mu}=0$ we increased the range of $M$ to 4096, which still
yielded a poor match. What we did note was that even for these sizable
values of $M$, convergence for $\tilde{\mu}=0$ is slow. When fitting the
data for $\tilde{\mu}=0$ for the ground state to
Eq.~\ref{FittedEquation} we are trying to force it to fit a
coefficient to a $1/M$ term which is not supposed to
be there. It is more appropriate to use the form
\begin{eqnarray}
aE = \ln(u_0) + {c_1\over M^{\beta}} 
\exp\left({c_2\over M}+{c_3\over M^2}\right),
\label{FittedEquation2}
\end{eqnarray}
where the power $\beta$ of the leading behavior is fitted
dynamically. We performed this refined fit to the three lowest lying 
states for $\tilde{\mu}=0$ which yielded the results assembled in
Table~\ref{fits2}.  
\begin{table}[h]
\begin{center}
\begin{tabular}{|c|c|}
\hline
& \hphantom{this}$\beta$\hphantom{this} \\
\hline
{\rm ground state} 	& 1.108 \\
{\rm 1st} 		& 1.025 \\
{\rm 2nd} 		& 1.015 \\
\hline
\end{tabular}
\end{center}
\caption[]{Table of fits to $\beta$ in Eq.~\ref{FittedEquation2} for
$\tilde{\mu}=0$.}
\label{fits2}
\end{table}
These results provide numerical evidence that for $\tilde{\mu}=0$, the 1st
and 2nd excited states do have a nonzero meson mass (i.e. the leading
behavior is $1/M$). However, the leading behavior for the ground state
decreases more rapidly than $1/M$ and is consistent with zero 
meson mass. 

\begin{figure}[ht]
\centerline{\psfig{file=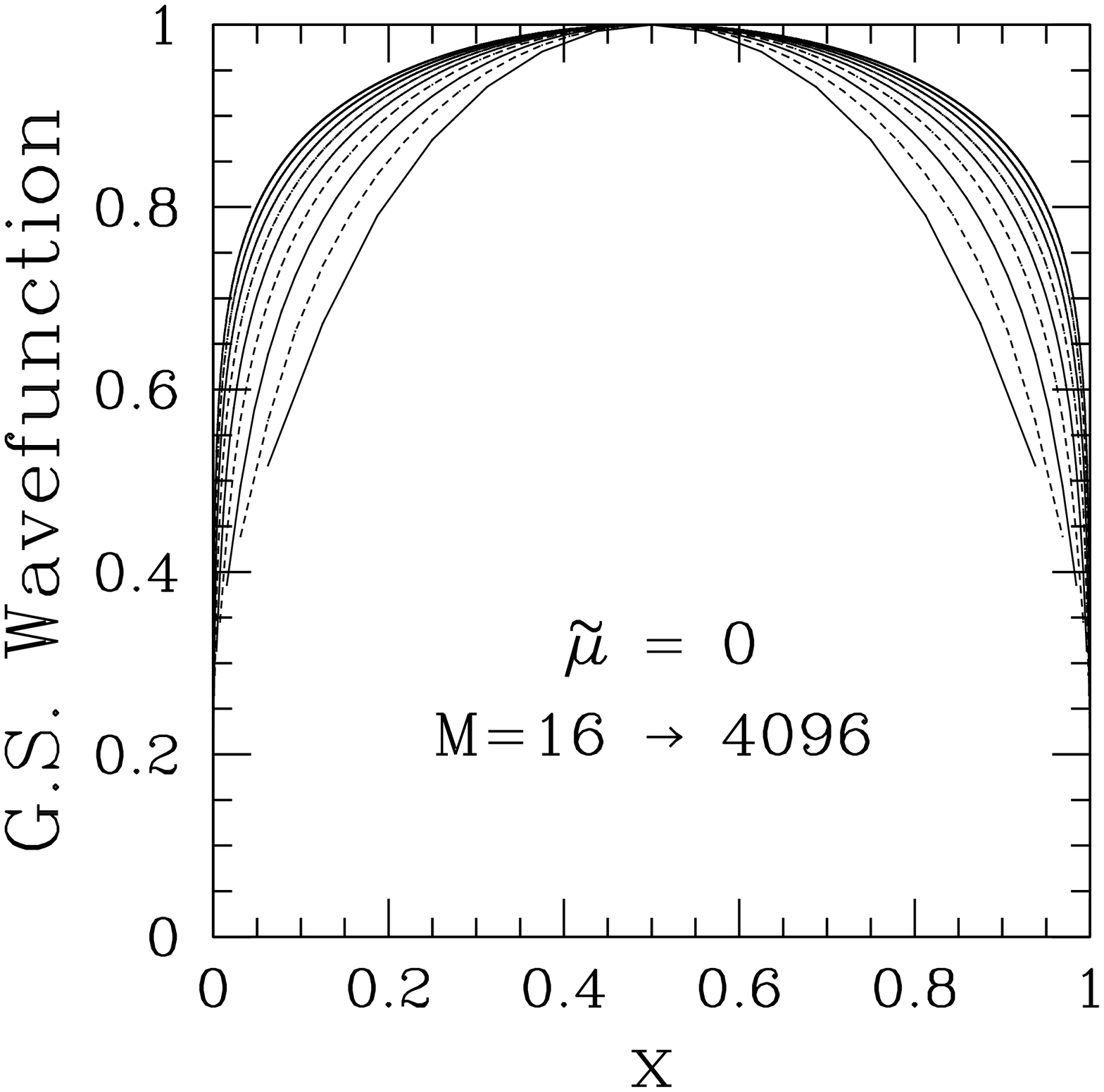,height=2.9in}\qquad
\psfig{file=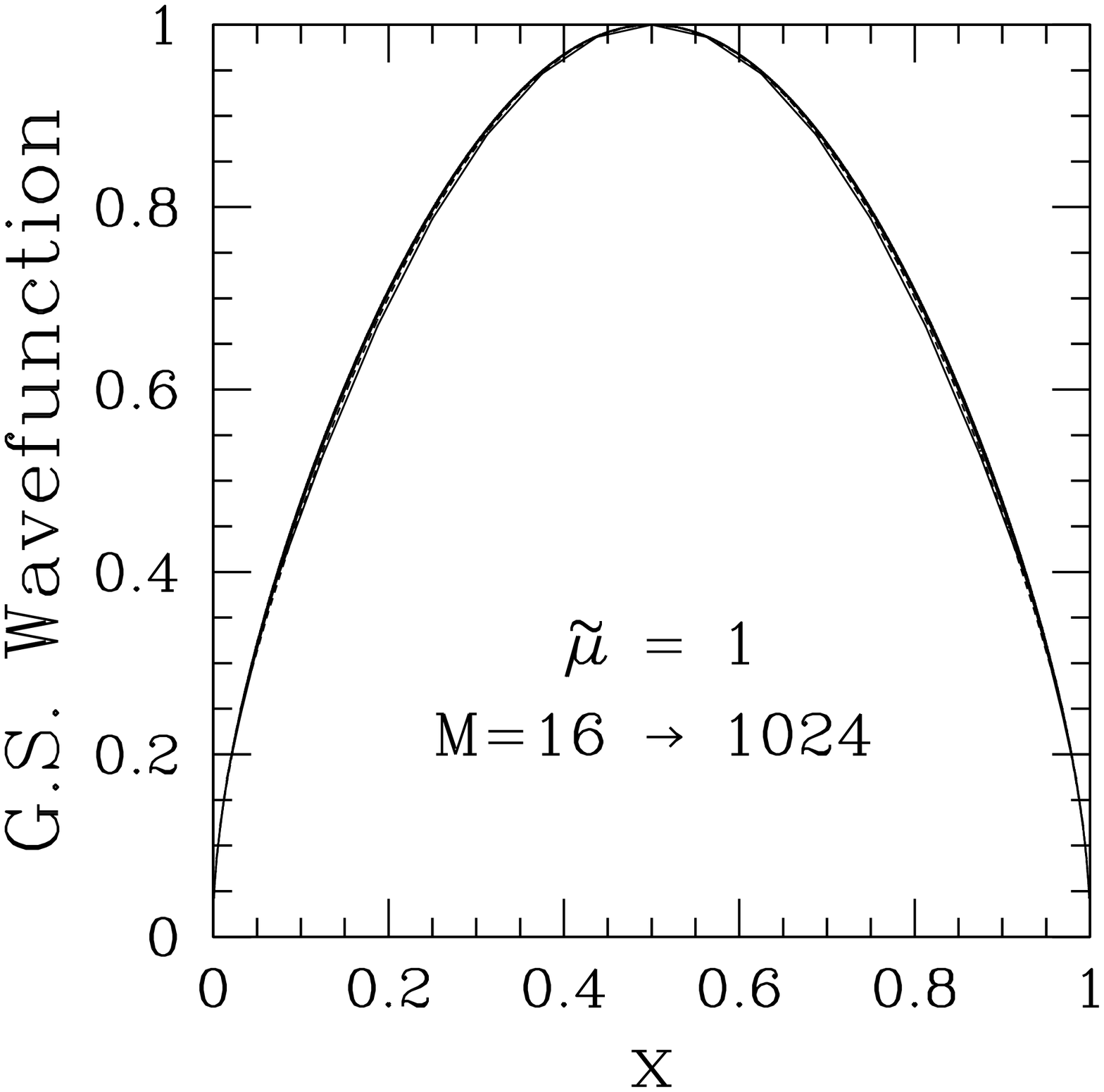,height=2.9in}}
\caption[]{Plot of the ground state eigenvector against $x=l/M$ 
for increasing $M$ for the cases $\tilde{\mu}=0$ and $1$. Each 
eigenvector is plotted for the range of
$M$ indicated in powers of $2$.}
\label{eigenvectors}
\end{figure}

We next address the issue of slow convergence for $\tilde{\mu}=0$ 
by examining the form of the ground energy eigenvector
for increasing values of $M$. It is well known that
the solutions of Eq.~\ref{thoofteq} for $\mu=0$ do not vanish
at the endpoints $x=0,1$; indeed the exact ground state is simply a
constant. As we can see in
Fig.~\ref{eigenvectors}, at finite large $M$ the
ground state solution of our discretized equation is ever
smaller at the endpoints, and the progression
of shapes is toward a more square profile. 
But even for $M=4096$ the eigenvector has not yet
converged to its limiting form. This should be compared with the
solution for $\tilde{\mu}=1$ which rapidly approaches it's limiting
form (see r.h.s. of Fig.~\ref{eigenvectors}).  
We see that, for our discretized equation, the  solution
for the ground state decreases more rapidly near the endpoints ($x=0$ and
$x=1$) as $M$ increases, consistently with the shape
eventually approaching a square profile at $M\to\infty$. 
However, it is not hard to show that consistency
of the continuum limit requires that the range in $x$ over which the fall-off
occurs must decrease less rapidly than $1/\sqrt{M}$. This
still allows an approach to a square profile but  
convergence is necessarily slower than one might
have expected.  In fact all solutions of the
continuum 't Hooft equation with $\tilde{\mu}=0$
have non-zero values at the end points. 
Thus we should expect slow convergence for all solutions
of the ${\tilde\mu}=0$ equation because the
discrete solution tends to vanish at the endpoints but the limiting
form does not. This
effect does not occur for $\tilde{\mu}>0$ because then
the continuum solution vanishes at the endpoints, so
a decent approximation to it can be achieved with relatively
smaller $M$.

\setcounter{equation}0
\renewcommand{\theequation}{\thesection.\arabic{equation}}
\section{Discussion and Conclusion}
\label{conc}
In this paper we have explored the efficacy of the discretization
of large $N_c$ QCD proposed in~\cite{beringrt} by applying it
to the well-understood 't~Hooft model. For a smooth continuum limit
over the whole range of bare coupling $\kappa$, we had to introduce
a refinement of the discrete time gluon emission
vertex. This amounted to insisting that after an emission, at
least 2 time steps had to intervene before the next emission,
with a similar restriction on consecutive absorptions. In contrast,
an absorption is allowed to immediately follow
an emission and {\it vice versa}. With this refinement in
place we found that the continuum 't~Hooft equation describes
the mass spectrum for all real $\kappa$. However, the parameters
that occur in the equation are renormalized from their bare
values, as summarized in 
Eqs.~\ref{ukappaeta},~\ref{effcoupling},~\ref{effmass}.

An amusing outcome of this renormalization phenomenon is that
the effective coupling goes to zero in both the small and
large $\kappa$ limits. Perhaps this feature is a version
of weak/strong coupling duality, much celebrated in
recent developments in string/M theory. 
However, we must concede that
2 dimensional QCD may be too trivial to expect anything other
than the usual continuum theory to emerge from any continuum
limit. Another caveat against
attributing much significance to this ``duality'' phenomenon, is that
the physics of the continuum limit really only depends
on the ratio $\mu^2/N_c g^2$. This is because one can
always choose the effective coupling as the fundamental
unit of energy. Then the theories at different coupling but
with the same value of this ratio (0 for example) are physically identical:
any differences in description can be removed by a change of units.  

At any rate, we conclude that the discretization of~\cite{beringrt}
can be meaningfully applied to QCD in 2 space-time dimensions, with
some intriguing hints about the nature of weak/strong coupling duality.
An obvious and important limitation of the 2 dimensional case, however,
is that the gluon has no dynamical degrees of freedom. Thus there
is no opportunity for the $P^+$ of the system to be shared amongst
an infinite number of gluons. This must occur for the fishnet
diagrams to be relevant, and is allowed in higher dimensional
space-time. The next step is to study the three dimensional case,
the simplest gauge theory where fishnet diagrams can be relevant.

\underline{Acknowledgements:} We thank Klaus Bering for
his helpful contributions in the early stages of this project.

\end{document}